\documentclass[12pt]{article}
\usepackage{graphicx}
\usepackage{url} 
\usepackage{amsmath}
\usepackage{amsthm}
\usepackage{amsfonts}
\usepackage{hyperref}
\usepackage{natbib} 
\usepackage{bm}
\usepackage{subcaption} %
\usepackage{multirow}  
\usepackage{booktabs}  
\usepackage{array}     
\usepackage{color}
\usepackage[table]{xcolor} 
\usepackage{colortbl} 
\usepackage{multirow} 
\usepackage{csquotes}
\usepackage{algorithm}
\usepackage{algpseudocode} 
\newcommand{\blind}{0}

\begin{document}

\def\spacingset#1{\renewcommand{\baselinestretch}%
{#1}\small\normalsize} \spacingset{1}


\if0\blind
{
  \title{\bf A new framework for non-stationary spatio-temporal data fusion of multi-fidelity models}
\author{Pietro Colombo\\
    School of Mathematics and Statistics, University of Glasgow\\
    University Place, Glasgow G12 8QQ, UK\\
    \texttt{pietro.colombo@glasgow.ac.uk}
    \and
    Fabio Sigrist\\
    Seminar for Statistics at ETH Zürich\\
    Rämistrasse 101, 8092 Zürich, Switzerland\\
    \texttt{fabio.sigrist@math.ethz.ch}
   \and
    Claire Miller\\
    School of Mathematics and Statistics, University of Glasgow\\
    University Place, Glasgow G12 8QQ, UK\\
   \texttt{claire.miller@glasgow.ac.uk}
    \and
    Ruth O’Donnell\\
    School of Mathematics and Statistics, University of Glasgow\\
    University Place, Glasgow G12 8QQ, UK\\
   \texttt{ruth.haggarty@glasgow.ac.uk}
    \and
    Xiaochen Yang\\
    School of Mathematics and Statistics, University of Glasgow\\
   University Place, Glasgow G12 8QQ, UK\\
    \texttt{xiaochen.yang@glasgow.ac.uk}
    \and
     Paolo Maranzano\\
    Department of Economics, Management and Statistics\\
   University of Milano-Bicocca, Piazza dell'Ateneo Nuovo, 1 - 20126, Milan, Italy\\
   Fondazione Eni Enrico Mattei, Corso Magenta , 63 - 20124, Milan, Italy\\
    \texttt{paolo.maranzano@unimib.it}
}

  \maketitle
} \fi

\if1\blind
{
  \bigskip
  \bigskip
  \bigskip
  \begin{center}
    {\LARGE\bf Title}
\end{center}
  \medskip
} \fi

\bigskip
\begin{abstract}
We propose a new scalable framework for spatio-temporal data fusion with multi-fidelity Gaussian processes (MFGPs) that enables fully likelihood-based inference for both stationary and non-stationary fidelity integration. The framework is designed for environmental applications, where abundant but noisy low-fidelity data (e.g., satellite or reanalysis products) must be fused with sparse yet accurate high-fidelity in-situ observations to obtain high-resolution reconstructions. Our key methodological contribution is a decomposed multi-fidelity covariance formulation that allows the Vecchia approximation to be applied directly to the latent low-fidelity and discrepancy processes. Combined with a Woodbury-based reconstruction, this yields a numerically stable and computationally efficient evaluation of the joint marginal likelihood without ever forming the full multi-fidelity covariance matrix. In addition, we introduce a generalized least squares (GLS) mean-removal strategy with fidelity-specific offsets, preventing systematic biases from being absorbed into cross-fidelity dependence. We validate the proposed approach through extensive experiments on synthetic data and a large-scale real-world application to wind speed reconstruction in the Lombardy region of Italy. The results show that the proposed Vecchia-based MFGP closely matches exact multi-fidelity inference in controlled settings, while substantially outperforming standard single-fidelity spatio-temporal Gaussian processes in terms of predictive accuracy, correlation, and representation of local variability in realistic large-data scenarios.
\end{abstract}

\noindent%
\vfill

\newpage
\spacingset{1.8} 
\section{Introduction}
\label{sec:intro}
Multi-fidelity models are a data-fusion approach based on Gaussian processes (GPs). They represent a promising solution for maximising the utility of multiple data sources 
(see \citealp{FernandezGodino2023,brevault2020overview}), delivering high-quality predictions.

\noindent
\hangindent=-5mm
There is extensive research on multi-fidelity models, addressing a wide range of challenges in fields such as engineering (i.e., \citet{toal2011efficient}, \cite{peherstorfer2019multifidelity}, \cite{kaps2022hierarchical}) and environmental science (\cite{colombo2025warped}, \cite{babaee2020multifidelity}). Early work, such as \citet{Kennedy2000}, laid the foundation, while more advanced approaches, including NARGP in \citet{perdikaris2017nonlinear} and deep Gaussian processes in \citet{damianou2013deep}, pushed the field forward. A drawback of this class of models, as with the Gaussian process framework in general, is that their computational cost is prohibitive for large datasets.

\noindent
\hangindent=-5mm
To address this challenge in the context of multi-fidelity modelling, various approaches have been proposed. For instance, \citet{cheng2021hierarchical} developed an autoregressive co-kriging framework by embedding Nearest-Neighbour Gaussian Process (NNGP) priors at each fidelity level. NNGP approximates the full Gaussian process using sparse precision matrices based on local (nearest-neighbour) interactions. However, it does not support marginal-likelihood computation, is limited to fully Bayesian models, and relies on a computationally intensive iterative MCMC procedure. More recently, \citet{cheng2024recursive} revived the recursive co-kriging approach of \citet{le2014recursive} to reduce the computational cost associated with full MCMC estimation.

\noindent
\hangindent=-5mm
We present here, to the best of our knowledge, the first integration of Vecchia approximations into a multi-fidelity Gaussian-process framework for likelihood-based inference. We also extend the spatio-temporal multi-fidelity model of \citet{babaee2020multifidelity} by introducing spatially varying cross-fidelity parameters—an innovation for spatio-temporal data applications. Third, we propose a generalized least squares estimation that allows mean level adjustment to the HF and LF signals enabling the GP to focus on modelling the shared spatio-temporal structure rather than the structural offset.

\noindent
\hangindent=-5mm
The use of the Vecchia approximation, applied to models that can incorporate a spatially varying cross-fidelity parameter, offers a new framework for implementing multi-fidelity models. This approach has proven to be not only computationally more efficient but also more numerically stable due to the regularization introduced by Vecchia approximation. This is primarily due to the sparsity of the Cholesky factor obtained via the Vecchia approximation. The high proportion of zero entries significantly reduces the risk of rounding errors during multiplication with dense vectors (see Appendix \ref{PositiveSemiDefinite} for further details).

The framework is particularly relevant in environmental studies, where both low-quality (low-fidelity, LF) and high-quality (high-fidelity, HF) data naturally occur. For example, satellite retrievals provide large geographic coverage when representing environmental variables such as wind speed, temperature, and humidity, but their representation of the phenomena of interest tends to be measured with a high level of uncertainty. By contrast, in-situ environmental networks provide highly accurate measurements of environmental variables, but data collection is restricted to specific station locations (i.e., in-situ monitoring) due to the high cost of equipment, logistical constraints, and accessibility challenges in certain terrains.

\noindent
\hangindent=-5mm
A key advantage of using a multi-fidelity framework for environmental-variable prediction is that it relies solely on spatio-temporal information. In other words, we can reconstruct environmental variables with higher accuracy without the need for external regressors (although these could, in principle, also be included in the framework).

\noindent
\hangindent=-5mm
The remainder of this work is organised as follows. Section 2 introduces the fundamental mathematical tools that underpin our framework. We begin by presenting the linear multi-fidelity Gaussian process model, then explain the implementation of the Vecchia approximation, and finally illustrate how the framework supports a non-stationary (spatially varying) integration of fidelity levels—or, more generally, any nonlinear connection between the levels. Section 3 demonstrates the practical effectiveness of the proposed framework, comparing the approximated models with the non-approximated version, by applying to large spatio-temporal datasets under controlled training and testing scenarios, with both synthetic and real-data experiments. Section 4 discusses the practical aspects and limitations concerning scalability, as well as possible further extensions. Finally, Section 5 concludes the paper with some closing remarks.

\section{Method}
\subsection{Background} \label{MFmodel}
Multi-fidelity models represent a multivariate extension of Gaussian process regression. The first formulation of such models appeared with the work of \cite{Kennedy2000}, which introduced the concept of using two Gaussian processes: one to model LF data and another to capture the discrepancy between HF and LF data. The HF response was expressed as a linear combination of the LF process and the discrepancy process.

The model can be described by the following equation:
\begin{equation}
    f_H(x) = \rho f_L(x) + \delta(x),
\end{equation}
where: \( f_L(x) \sim GP(\mu(x), k(x,x')) \) represents the LF Gaussian process (i.e., a Gaussian process trained on data of lower accuracy), with mean function \( \mu(\cdot) \) and covariance function \( k(\cdot,\cdot) \). The residual discrepancies between the HF and LF data are modelled through another Gaussian process \( \delta(x) \sim GP(\mu_{\delta}(x), k_{\delta}(x,x')) \), independent of the first Gaussian process. The parameter \( \rho \) serves two purposes: it models the cross-covariance between the HF and LF Gaussian processes and rescales the LF data to match the signal variance of the HF data. 

\noindent 
\hangindent=-5mm
Broadly speaking, $f_H$ is also a Gaussian process; however, it is constructed by building upon the $\delta$ and $f_L$ processes. Given some training data \( [\bm{y}_L, \bm{y}_H] \) observed at input locations \( [\bm{x}_L, \bm{x}_H] \)\footnote{Notice that for a spatio-temporal model \( \bm{x_L} \) is a matrix with 3 columns and $n_L$ rows.}, where $\bm{y}_L$ is a vector of LF observations while $\bm{y}_H$ is a vector of HF observations, we model the observed LF data, \( y_L(x) \), and HF data, \( y_H(x) \), as noisy realizations of the true processes \( f_L(x) \) and \( f_H(x) \), respectively. For notational convenience, we use $x$ to denote a generic input location; in cases where HF and LF data are aligned, 
$x$ also represents the matched location across fidelities. In particular, we assume:
\begin{align}
    y_L(x) &= f_L(x) + \varepsilon_L, \quad \varepsilon_L \overset{\text{i.i.d.}} \sim \mathcal{N}(0, g^2_L), \\
    y_H(x)&= \rho\, f_L(x) + \delta(x) + \varepsilon_{\delta}, \quad \varepsilon_{\delta} \overset{\text{i.i.d.}} \sim \mathcal{N}(0, g^2_{\delta}).
\end{align}
The vector $\bm{y}=[\bm{y}_L, \bm{y}_H]^\top$ has a multivariate normal distribution:
\begin{equation}
    \bm{y}\equiv\begin{bmatrix}
        \bm{y}_L\\ \bm{y}_H \\ 
    \end{bmatrix}~\sim \mathcal{N}_n\big(\bm{\mu}, \bm{K} \big),
\end{equation}
which is defined by the covariance matrix $\bm{K}$ and the mean vector $\bm{\mu}$. The entries of $\bm{K}$ depend on the chosen covariance function, such as the RBF kernel:
\[
k(x,x') = \sigma_s\exp\left(-\frac{\|x - x'\|^2}{2 l^2}\right).
\]
In this kernel, $\sigma_s$ is the signal variance parameter, and $l$ is known as the \textit{length-scale} or \textit{decay} parameter, which defines the speed at which correlation decays with distance.
In principle, any kernel could be used in this framework. We selected this kernel due to its suitability for the application discussed later. 
The structure of the multi-fidelity (MF) covariance matrix $\bm{K}$ is defined as follows:

\begin{equation}
\bm{K} = \begin{bmatrix}
\bm{K}_{LL} & \bm{K}_{LH} \\
\bm{K}_{HL} & \bm{K}_{HH}
\end{bmatrix} = 
\begin{bmatrix}
k_{LL}(\bm{x}_L, \bm{x}_L; \theta_L) + g_L^2 \bm{I} & k_{LH}(\bm{x}_L, \bm{x}_H; \theta_L, \rho)\\
k_{HL}(\bm{x}_H, \bm{x}_L; \theta_L, \rho) & k_{HH}(\bm{x}_H, \bm{x}_H; \theta_L, \theta_{\delta}, \rho) + g_{\delta}^2 \bm{I}
\end{bmatrix}
\end{equation}

\noindent \textbf{Dimensions and Shapes:}
\begin{itemize}
    \item $n_L, n_H$: Number of low- and high-fidelity spatial/temporal points, respectively.
    \item $\bm{K}_{LL} \in \mathbb{R}^{n_L \times n_L}$: Square covariance matrix for low-fidelity data.
    \item $\bm{K}_{HH} \in \mathbb{R}^{n_H \times n_H}$: Square covariance matrix for high-fidelity data.
    \item $\bm{K}_{LH} \in \mathbb{R}^{n_L \times n_H}$: Cross-covariance block linking the two fidelities.
    \item $\bm{K}_{HL} \in \mathbb{R}^{n_H \times n_L}$: Transpose of the cross-covariance block ($\bm{K}_{LH}^T$).
    \item $\bm{K} \in \mathbb{R}^{(n_L+n_H) \times (n_L+n_H)}$: The full joint covariance matrix.
\end{itemize}
\noindent 
\hangindent=-5mm
Notice that $\theta=[\sigma_s,l]$, and we use the subscript $L$ or $\delta$ to distinguish between the processes, i.e., $\theta_L$ are the parameters of the covariance of $f_L$, while $\theta_{\delta}$ are the parameters of the covariance of $\delta$. $g_\delta^2$ and $g_L^2$ are the nugget variances of the two Gaussian processes. The terms $k_{LH}$ and $k_{HH}$\footnote{Note that we are using the generic input notation, $x,x'$ rather than the MF-specific $x_L$, $x_H$.} are given by: 
\[
    k_{HL}(x, x'; \theta_L, \rho) = k_{LH}(x, x'; \theta_L, \rho) = \rho k_1(x, x'; \theta_L),
\]
and 
\[
    k_{HH}(x, x'; \theta_L, \theta_{\delta}, \rho) = \rho^2 k_1(x, x'; \theta_L) + k_2(x, x'; \theta_{\delta}).
\]
\noindent
\hangindent=-5mm
This model for $x \in \mathcal{R}$ has 7 parameters $[g_L, g_{\delta}, \sigma^L_s,\sigma^{\delta}_s,l_L,l_{\delta}, \rho]$: three for each Gaussian process plus $\rho$. The parameters can be trained by minimising the negative marginal log-likelihood:
\begin{equation}
    \mathcal{NLML}(\theta_L, \theta_{\delta}, \rho) = \frac12 \bm{y}^\top \bm{K}^{-1} \bm{y} + \frac12 \log |\bm{K}| + \frac{n}{2} \log(2 \pi).
\end{equation}

\noindent
\hangindent=-5mm
Note that, depending on the choice of covariance function and the number of input dimensions, the number of parameters can grow substantially. The 7 parameters apply only to the one-dimensional case with the RBF kernel.

\subsubsection{The Vecchia approximation}
The Vecchia approximation \citep{vecchia1988estimation, datta2016hierarchical,katzfuss2021general} is one of the most popular and effective \citep{rambelli2025accuracy} methods used in spatial (spatio-temporal) statistics to approximate the precision matrix of a Gaussian process. It is applied to a myriad of contexts such as multi-scale modelling \citep{zhang2022multi}, or in combination with the Laplace approximation to model a latent Gaussian process  \citep{kundig2024iterative}.  Such a class of approximations is an alternative to the sparse Gaussian Process framework, e.g., \cite{lalchand2022sparse}, where inducing points are used for providing a global approximation of the models. While sparse GP works on low-rank approximations of the covariance, the Vecchia approximation introduces sparsity in the precision matrix. Given a fixed computational budget, the Vecchia approximation framework provides greater accuracy for spatial data, as demonstrated by \cite{rambelli2025accuracy}.

Consider a Gaussian vector \[
\bm{y} = \begin{pmatrix}
y_1 \\
y_2 \\
\vdots \\
y_n
\end{pmatrix} \sim \mathcal{N}_n(\bm{0}, \bm{K}),
\]
where $\bm{K}$ is a $n \times  n$ positive semidefinite covariance matrix. The minimisation of the likelihood generally requires the inversion of $\bm{K}$, which has $O(n^3)$ time complexity and $O(n^2)$ memory storage. The Vecchia approximation implies that each \( y_i \) is conditionally independent of all previous observations not in a neighborhood set \( C(i) \), given \( y_{C(i)} \). This assumption allows the likelihood to be computed in \( \mathcal{O}(nm^3) \) time, where \( m \) is the maximum size of the conditioning set  \( C(i) \), making the computation of the precision matrix feasible to be calculated iteratively. In essence, the idea is to approximate the full dependency structure using only a small set of local neighbors, under the assumption that the rest contribute little additional information. In particular, the exact factorization of a Gaussian density is \[
p(y) = p(y_1) p(y_2 \mid y_1) p(y_3 \mid y_1, y_2) \cdots p(y_n \mid y_1, y_2, \ldots, y_{n-1}),
\] and hence in the Vecchia approximation the joint probability can be written as \[
p(y) \approx \prod_{i=1}^n p(y_i \mid y_{C(i)}).
\]
 The maximum size $m$  of the conditioning set is decided by the user and is assumed to be much smaller than the sample size $n$. The smaller the value of $m$, the greater the computational gain, but the poorer the approximation. The approximate covariance matrix $\hat{\bm{K}}$ has a sparse inverse Cholesky factor
\[
\hat{\bm{K}}^{-1} = \bm{U} \bm{U}^\top, 
\]
where $\bm{U}$ is a sparse upper triangular matrix. When the neighbour size $m$= $n-1$, the approximation is exact.

\subsection{A new framework for the estimation of scalable MFGP}
\label{NewMethod}
Multi-fidelity models are, in essence, hierarchical models in which independent Gaussian
processes are used to represent data of varying quality levels. In this section, we introduce a new framework for implementing MFGPs that is scalable, fully likelihood-based, and numerically stable. The proposed method integrates three key components: (1) a multi-fidelity GP formulation, (2) the Vecchia approximation for scalability, and (3) an efficient decomposition and reconstruction of the covariance structure. 


\noindent
\hangindent=-5mm
The main challenge is that the implementation of the Vecchia approximation is difficult to achieve directly for $\bm{K}$, since it is unclear how the conditioning sets should be chosen, as it contains the cross-correlation between different processes ($k_{HL}$ and $k_{LH}$). However, we can rewrite the covariance model by separating the contribution of the discrepancy model from the LF model as follows:
\[
\bm{K} = \bm{A} \bm{\Sigma}_w \bm{A}^\top + \bm{D}_{\epsilon},
\]
where:
\begin{itemize}
    \item \( \bm{A} \) is a matrix that connects the LF and HF components.
    \item \( \bm{\Sigma}_w \) is a block-diagonal covariance matrix for the two Gaussian processes \( f_L \) and \( \delta \).
    \item \( \bm{D}_{\epsilon} \) is a diagonal matrix that represents the nugget error terms \( g_L \bm{I} \) and \( g_{\delta} \bm{I} \).
\end{itemize}

\noindent
\hangindent=-5mm
More precisely, the matrix \( \bm{A} \) defines how the LF and HF processes interact. Thus, \( \bm{A} \) is defined as:

\[
\bm{A} = \begin{bmatrix} \bm{Z}_1 & 0 \\ \rho \bm{Z}_{21} & \bm{I} \end{bmatrix}.
\]

\noindent
\hangindent=-5mm
Here, \( \bm{Z}_1 \) and \( \bm{Z}_{21} \) track the positions of the LF and HF data, respectively, and \( \rho \) scales these interactions. In this covariance decomposition, \( \bm{\Sigma}_w \) represents the covariance matrix of \( f_L \) and \( \delta \), defined as:
\[
\bm{\Sigma}_w = 
\begin{bmatrix} 
\bm{\Sigma}_L & 0 \\ 
0 & \bm{\Sigma}_{\delta} 
\end{bmatrix},
\]
\noindent
\hangindent=-5mm
where \( \bm{\Sigma}_L \) and \( \bm{\Sigma}_{\delta} \) are the covariance matrices corresponding to the LF and discrepancy Gaussian processes, respectively. As \( \bm{\Sigma}_L \) and \( \bm{\Sigma}_{\delta} \) follow standard Gaussian process covariance structures, the Vecchia approximation can be applied independently to each process, where  \( \bm{\Sigma}^{-1}_{L}=\bm{U}_L \bm{U}^\top_L \) and \( \bm{\Sigma}^{-1}_{\delta}= \bm{U}_{\delta} \bm{U}^\top_{\delta} \).

\noindent
\hangindent=-5mm
This decomposition offers several advantages. First, it allows the approximation to be implemented with differing levels of accuracy for the LF and HF data. Such flexibility is particularly valuable in multi-fidelity applications, where HF data are typically sparse whilst LF data are more abundant. In these circumstances, it is often practical to approximate only the LF data, making fuller use of the HF data. Second, a non-linear structure may readily be imposed on $\rho$, as discussed in Section~\ref{Non Stationary}. As the Vecchia approximation induces sparsity in the precision matrix, we additionally observe improved numerical stability.

The next challenge is to reconstruct the precision matrix $\bm{K}^{-1}$ of the full model. This can be done by employing a well-known matrix identity, the Woodbury identity. In particular, given that $\bm{\Sigma}^{-1}_w$ is defined as:
\[
\bm{\Sigma}^{-1}_w = 
\begin{bmatrix} 
\bm{\Sigma}^{-1}_{L} & 0 \\ 
0 & \bm{\Sigma}^{-1}_{\delta} 
\end{bmatrix},
\]
the inverse of the full covariance can be expressed as:
\[
\bm{K}^{-1} = \bm{D}_{\epsilon}^{-1} - \bm{D}_{\epsilon}^{-1} \bm{A} \left( \boldsymbol{\Sigma}_{w}^{-1} + \bm{A}^\top \bm{D}_{\epsilon}^{-1} \bm{A} \right)^{-1} \bm{A}^\mathsf{T} \bm{D}_{\epsilon}^{-1}.
\]

\noindent
\hangindent=-5mm
In this formulation, we define
\[
\bm{H} = \boldsymbol{\Sigma}_{w}^{-1} + \bm{A}^\top \bm{D}_{\epsilon}^{-1} \bm{A},
\]
where $\bm{H}$ is sparse, and all linear solves
are performed using sparse Cholesky factorization with a fill-reducing
approximate minimum degree (AMD) permutation.
The computational efficiency of our likelihood evaluation stems from the sparsity of the auxiliary precision matrix
\[
\bm{H}=\bm{\Sigma}_w^{-1}+\bm{A}^\top \bm{D}_\epsilon^{-1}\bm{A},
\]
whose sparsity is primarily induced by the Vecchia approximation to $\bm{\Sigma}_w^{-1}$ (with an additional sparse contribution from $\bm{A}^\top \bm{D}_\epsilon^{-1}\bm{A}$). Since the precision of the Vecchia conditioning depends on the chosen ordering, the effect of ordering on the sparsity structure of $\bm{H}$ is investigated in Appendix~\ref{sec:supp_ordering_H} (see also Figure~\ref{fig:H_ordering} and Table~\ref{tab:H_ordering}).

The inverse of $\bm{K}$ is never formed explicitly. Instead, linear solves with $\bm{K}$ (and hence likelihood evaluation) are carried out via the Woodbury identity, which requires constructing the Vecchia factors for the LF and discrepancy processes and factorizing $\bm{H}$.  

Before evaluating the marginal likelihood, we remove fidelity-specific mean offsets using a generalised least squares (GLS) procedure. 
Specifically, we allow the low- and high-fidelity observations to have separate intercept terms, which are estimated under the joint covariance $\bm{K}$ implied by the multi-fidelity Gaussian process model. 
All likelihood and prediction computations are subsequently performed using the 
centred residual vector $\tilde{\bm{y}} = \bm{y} - \bm{G}\hat{\bm{\beta}}$, 
where $\bm{G}$ is the GLS design matrix for the mean structure and 
$\hat{\bm{\beta}}$ contains the estimated mean coefficients. This approach 
ensures that the covariance structure explicitly models spatio-temporal 
dependence, rather than being confounded by systematic baseline differences 
between fidelities. Further details of the GLS formulation and its spatially 
dependent case are provided in Appendix~\ref{app:gls_mean}.

The resulting dominant computational cost can be summarized as
\[
\text{Total Cost}
=
\mathcal{O}\!\left(n_L m_L^2\right)
+
\mathcal{O}\!\left(n_\delta m_\delta^2\right)
+
\mathcal{O}\!\left(\mathrm{nnz}(R)\right),
\]
where $n_L$ and $n_\delta$ denote the sizes of the LF and discrepancy latent vectors (i.e., the dimensions of $\boldsymbol{\Sigma}_L$ and $\boldsymbol{\Sigma}_\delta$), $m_L$ and $m_\delta$ are the corresponding Vecchia neighborhood sizes, and $R$ is the sparse Cholesky factor of a fill-reduced permutation of $\bm{H}$ (so $\mathrm{nnz}(R)$ quantifies the factorization complexity through fill-in). This is substantially more efficient than direct factorization of $\bm{K}$, which scales as $\mathcal{O}(n^3)$ with $n=n_L+n_\delta$. A computation-time comparison between the standard and our likelihood implementation is reported in Appendix~\ref{fig:computation_time}.

We now present the proposed MFGP model in matrix form:
\begin{equation}
\begin{split}
\bm{y}_L &= \bm{Z}_1 \bm{w}_L + \bm{\varepsilon}_L,
\qquad \bm{y}_L \in \mathbb{R}^{n_L},\ 
\bm{Z}_1 \in \{0,1\}^{n_L \times N_1},\ 
\bm{w}_L \sim GP\!\left(\bm{0}, \boldsymbol{\Sigma}_L\right), \\
\bm{y}_H &= \rho\, \bm{Z}_{21}\bm{w}_L + \bm{w}_\delta + \bm{\varepsilon}_\delta,
\qquad \bm{y}_H \in \mathbb{R}^{n_\delta},\ 
\bm{Z}_{21}\in \{0,1\}^{n_\delta \times N_1},\ 
\bm{w}_\delta \sim GP\!\left(\bm{0}, \boldsymbol{\Sigma}_\delta\right).
\end{split}
\end{equation}

\noindent

The only term not previously defined is $N_1$, which denotes the number of distinct spatial locations.
In a nested design, every high-fidelity location is also observed at low fidelity, so no HF-only locations exist and $N_1 = n_L$.
In a non-nested design, some high-fidelity observations may be available at locations where no low-fidelity data are observed.
In this case, $N_1$ is equal to the total number of unique spatial locations obtained by taking the union of locations where only $y_L$ is observed, locations where only $y_H$ is observed, and locations where both $y_L$ and $y_H$ are observed. See Algorithm 1 for a compact overview of the new vecchia approximated likelihood based procedure.

\clearpage   

\begin{algorithm}[t]
\caption{Vecchia-based MFGP likelihood with GLS mean removal}
\label{alg:vecchia_mfgp_gls}
\footnotesize
\begin{algorithmic}[1]
\Require $\boldsymbol{\theta}$, $(\bm{X}_L,\bm{y}_L)$, $(\bm{X}_H,\bm{y}_H)$
\Ensure $\mathcal{NLML}(\boldsymbol{\theta})$

\State $\bm{y}\leftarrow[\bm{y}_L^\top,\bm{y}_H^\top]^\top$
\State Build Vecchia factors $\boldsymbol{\Sigma}_L^{-1}$ and $\boldsymbol{\Sigma}_\delta^{-1}$
\State Compute $\log|\boldsymbol{\Sigma}_w|$ from Vecchia diagonals

\If{$\rho(\cdot)$ non-stationary}
    \State Estimate $\boldsymbol{\rho}_H$ via empirical scaling and GP smoothing
\Else
    \State $\boldsymbol{\rho}_H\leftarrow\rho\bm{1}$
\EndIf

\State Construct $\bm{A}$ and $\bm{D}_\epsilon$
\State $\bm{H}\leftarrow \boldsymbol{\Sigma}_w^{-1}+\bm{A}^\top\bm{D}_\epsilon^{-1}\bm{A}$
\State Sparse Cholesky: $\bm{P}\bm{H}\bm{P}^\top=\bm{R}^\top\bm{R}$
\State Compute $\log|\bm{H}|$

\State Build GLS design matrix $\bm{G}$
\State Solve $\bm{K}^{-1}[\bm{y},\bm{G}]$ via Woodbury
\State $\hat{\boldsymbol{\beta}}\leftarrow(\bm{G}^\top\bm{K}^{-1}\bm{G})^{-1}\bm{G}^\top\bm{K}^{-1}\bm{y}$
\State $\tilde{\bm{y}}\leftarrow\bm{y}-\bm{G}\hat{\boldsymbol{\beta}}$

\If{if GLS == False}
\State $\mathcal{NLML}\leftarrow \frac12\tilde{\bm{y}}^\top\bm{K}^{-1}\tilde{\bm{y}}
+\frac12(\log|\boldsymbol{\Sigma}_w|+\log|\bm{H}|+\log|\bm{D}_\epsilon|)
+\frac{N}{2}\log(2\pi)$
\Else
\State $\mathcal{NLML}_{\mathrm{REML}}
=
\frac{1}{2}\,
\tilde{\bm y}^{\top}\bm K^{-1}\tilde{\bm y}
+
\frac{1}{2}
\Big(
\log|\boldsymbol{\Sigma}_w|
+
\log|\bm H|
+
\log|\bm D_{\epsilon}|
+
\log|\bm G^{\top}\bm K^{-1}\bm G|
\Big)
+
\frac{N-P}{2}\log(2\pi)$
\EndIf
\State \Return $\mathcal{NLML}$
\end{algorithmic}
\end{algorithm}

\clearpage   

\subsection{Spatial non-stationary variation}\label{Non Stationary}

The second contribution of our work is to propose a spatio-temporal model with a non-stationary integration of the different fidelity levels. In other words, we assume $\rho$ to be a function rather than a parameter. This assumption is not entirely new, as previous works such as \citet{perdikaris2017nonlinear} and \citet{raissi2016deep} have explored non-linear connections between fidelity levels. More recently, \cite{sauer2023vecchia} implemented a Vecchia-approximated deep Gaussian process model which, under specific assumptions and with multiple layers, can be interpreted as a complex multi-fidelity framework. However, we highlight here some key differences: our approach simplifies the implementation of a non-linear integration of fidelity levels. Implementing a non-linear $\rho(\cdot)$ without a defined sparse covariance structure can lead to the resulting covariance matrix being non-positive definite.The key idea is simply that the use of sparse matrices improves numerical stability, and hence positive semi-definiteness. More details are provided in Appendix \ref{PositiveSemiDefinite}. Second, the implementation of a non-linear integration function for spatio-temporal environmental studies is new. This extends previous work on spatio-temporal mapping and prediction, such as \cite{babaee2020multifidelity} and \cite{colombo2025warped}. Third, we introduce a specific input-coordinate non-linear integration function. This enhances the interpretability of the integration between fidelity levels and avoids introducing excessive wiggliness that could result from including additional dimensions.

\noindent
\hangindent=-5mm

Consider the following MF covariance matrix obtained with our framework:
 
\[
\bm{K} = 
\begin{bmatrix} 
\bm{Z}_1 \boldsymbol{\Sigma}_{L} \bm{Z}_1^\top & 
\rho\, \bm{Z}_1 \boldsymbol{\Sigma}_{L} \bm{Z}_{21}^\top \\ 
\rho\, \bm{Z}_{21} \boldsymbol{\Sigma}_{L} \bm{Z}_1^\top & 
\rho^2\, \bm{Z}_{21} \boldsymbol{\Sigma}_{L} \bm{Z}_{21}^\top + \boldsymbol{\Sigma}_{\delta} 
\end{bmatrix} 
+ \bm{D}_{\epsilon}.
\]

\noindent
\hangindent=-5mm
If we replace the parameter \( \rho \) with a function that depends on spatial coordinates, \( \rho(\bm{s}) \), the integration of different fidelity levels will vary with location. In theory, one could use a function that varies with both space and time (i.e., \( \rho(s,t) \)). 
\noindent
\hangindent=-5mm
Below, we present some examples of functions that we implemented; the choice of function is highly dependent on the application. In our implementation, we choose simpler functions that vary only in space, such as linear or polynomial forms. For example, if we have two spatial coordinates, \( s_1 \) and \( s_2 \) (with \( s = [s_1,s_2] \)), a linear function might be:
\begin{align}
    \rho(s_1,s_2) = \alpha_{\rho} + \beta^{s_1}_{\rho}\, s_1 + \beta^{s_2}_{\rho}\, s_2.
\end{align}
With this notation, $\alpha_{\rho}$ is the intercept term (or base level) of the function $\rho(s, s')$.
$\beta^{s_1}_{\rho}$ is the coefficient for the variable $s_1$ in the context of $\rho$. The superscript $s_1$ indicates the variable to which this coefficient belongs.
Similarly, the parameter $\beta^{s_2}_{\rho}$ is the coefficient for $s_2$ in $\rho$. We also test a second-order polynomial function, which includes the intercept, the first-order terms, and the squared terms:
\begin{align}
    \rho(s_1,s_2) = \alpha_{\rho} + \beta^{s_1}_{\rho}\, s_1 + \beta^{s_2}_{\rho}\, s_2 + \beta^{s^2_1}_{\rho}\, s_1^2 +  \beta^{s^2_2}_{\rho}\, s_2^2.
\end{align}
A third function that we test is the \textit{empirical Gaussian process} (eGP), in which the function $\rho(\cdot)$ is modelled as a smoothed version of its empirical estimates\footnote{This is simply a non-parametric smoother of the empirical field. No assumptions are made regarding the smoothing functions. In this context, the interaction between $s_1$ and $s_2$, the choice of kernels, and other considerations are less relevant and have little impact on the overall results.}. Specifically, for each spatial location, we compute the following empirical values:
\begin{equation}
    \rho(s_1, s_2) = \frac{\text{cov}(y_H, y_L)}{\text{var}(y_L)}.
\end{equation}
These values represent local empirical estimates of the linear relationship between the HF ($y_H$) and LF ($y_L$) data. Naturally, such estimates can only be computed at locations where both $y_H$ and $y_L$ are jointly available. Hence, for prediction locations, we use an interpolated $\hat{\rho}$. We then apply Gaussian process regression to these empirical values, modelling them as:
\begin{equation}
\rho(\bm{s}) = f(\bm{s}) + \epsilon_s,
\end{equation}
where $f(\bm{s})$ captures the spatially smooth underlying trend, and $\epsilon_s$ is a noise term. This function is used in our real data experiment, see Section \ref{Real_data_experiment}, since it shows the highest degree of flexibility. 

The function $\rho(\bm{s})$ should be interpreted as a location-dependent
regression slope linking the low-fidelity and high-fidelity signals.
In particular, at each spatial location where both fidelities are observed,
$\rho(\bm{s})$ corresponds to the local least-squares coefficient obtained
from regressing $y_H(\bm{s},t)$ on $y_L(\bm{s},t)$ over time,
that is,
\[
\hat{\rho}(\bm{s}) =
\frac{\widehat{\mathrm{Cov}}\!\left(y_H(\bm{s},\cdot),y_L(\bm{s},\cdot)\right)}
{\widehat{\mathrm{Var}}\!\left(y_L(\bm{s},\cdot)\right)}.
\]
This quantity measures the strength of the linear information transfer
from the low-fidelity to the high-fidelity process.
blueThe resulting $\rho(\bm{s})$ is treated as a deterministic, data-driven
function (after spatial smoothing), rather than as a latent stochastic
process jointly inferred within the multi-fidelity likelihood.  For completeness, we also implemented a fully stochastic version of the model, where the coupling parameter $\rho$ and the kernel covariance parameters are estimated jointly via maximum likelihood. However, this specification frequently encounters identifiability issues, as the likelihood function struggles to decouple the contribution of the covariance parameters in $\boldsymbol{\Sigma}_w$ from that of $\rho$. Furthermore, we observed that this fully stochastic formulation yields no measurable improvement in predictive accuracy compared to our primary approach. For completeness, such issues might strictly connected to the nature of our dataset, which present very few spatial locations.

Consequently, spatial variation in $\rho(\bm{s})$ induces
non-stationarity in the high-fidelity covariance structure through
the multiplicative term $\rho(\bm{s})\rho(\bm{s}')k_L(x,x')$.

The spatially varying rescaling is implemented
through a location-dependent function $\rho(s)$ evaluated at the HF
locations, which induces the HF--HF covariance block
\[
k_{HH}(x,x') = \rho(s)\rho(s')\,k_L(x,x') + k_\delta(x,x'),
\]
where $k_L$ denotes the LF covariance kernel and $k_\delta$ is the
independent discrepancy kernel; this form is symmetric and positive
semidefinite by construction.

For more information, about the positive semi-definiteness implied by such covariance see Appendix \ref{app:PSD_rho}. The inclusion of a flexible cross-fidelity function $\rho(\bm{s})$ increases the number of hyperparameters to be estimated, but it also enables the model to 
adapt to spatially heterogeneous LF–HF relationships. This formulation demonstrates 
that the proposed framework can accommodate a broad class of multi-fidelity models 
within a unified inference strategy.

%
%
%
%
%
%
%
%

\subsection{Kernels and spatio-temporal assumptions}
\label{sec:kernel_spacetime}

Let $x=x(\bm{s},t)$ denote a spatio-temporal input, with spatial coordinates $\bm{s}\in\mathbb{R}^d$ and time $t\in\mathbb{R}$. Both the low-fidelity latent process $f_L(x)$ and the discrepancy process $\delta(x)$ are modeled as zero-mean Gaussian processes with a separable (multiplicative) spatio-temporal covariance structure, such that for $\star\in\{L,\delta\}$ the covariance function satisfies $k_{\star}((\bm{s},t),(\bm{s}',t')) = k_{\star}^{(s)}(\bm{s},\bm{s}')\,k_{\star}^{(t)}(t,t')$. In this work, $k_{\star}^{(s)}$ and $k_{\star}^{(t)}$ are chosen as squared-exponential kernels with distinct spatial and temporal length-scale parameters for each process, allowing the low-fidelity and discrepancy components to exhibit different smoothness and correlation ranges. The separable formulation provides a pragmatic balance between modeling flexibility and computational scalability, is naturally compatible with the Vecchia approximation used for likelihood evaluation, and is consistently adopted in both the main methodology and the synthetic data generation described in Appendix~\ref{sec:sim_data}. While our MATLAB implementation supports a spatio-temporal separable Matérn kernel, empirical testing indicated that this covariance structure was less suitable for the characteristics of the current dataset.

\section{Experiments} \label{Experiments}
To demonstrate the effectiveness and reliability of the proposed method, we conduct three complementary experiments. In all experiments, Vecchia conditioning is performed using the native ordering of the data: observations are ordered by spatial location (space-major ordering), with temporal replicates ordered sequentially within each location. All experiments are performed by holding out entire target stations to simulate a total absence of HF information at test locations.

The first experiment is a targeted validation study designed to assess the numerical stability and accuracy of the decomposed Vecchia approximation under different  ordering strategies. The second experiment uses synthetic data to evaluate predictive performance under controlled conditions. The third experiment applies the method to real-world wind speed data to demonstrate its behaviour in a realistic, large-scale setting.

\noindent  
\hangindent=-5mm

\noindent  
\hangindent=-5mm  
The synthetic-data experiment evaluates model performance in a controlled setting, where the spatial locations are evenly distributed and the relationship between low- and high-fidelity data is strong and consistent. Under these conditions, local variability poses limited difficulty, and evaluation is more straightforward. We therefore use root mean squared error (RMSE) and mean absolute error (MAE) to assess prediction accuracy in magnitude. 

\noindent  
\hangindent=-5mm  
The real-data experiment presents additional challenges. The low-fidelity data provide only coarse approximations of the true high-fidelity measurements, with correlations ranging from 0.32 to 0.55, and the monitoring stations are unevenly distributed across space. In this setting, accurately capturing local variability is more important than minimising global error magnitude. Consequently, we replace RMSE with the correlation coefficient as a primary performance metric. Evaluation therefore relies on MAE to assess magnitude,  and the correlation coefficient to represent the ability of the model to reproduce local spatial and temporal patterns.

\subsection{Stability and accuracy of the decomposed Vecchia reconstruction}

The first validation experiment addresses the methodological integrity of the proposed decomposition strategy. By comparing the decomposed Vecchia approximation against an exact MFGP benchmark, we evaluate how approximation errors at the sub-component level affect the final Woodbury reconstruction. To isolate these approximation effects from optimization noise, we fixed the hyperparameters at the values obtained via exact inference and repeated the analysis over 20 independent data realizations ($n_\text{rep}$). The data are simulated using the simulation approach described in Appendix \ref{sec:sim_data}.
Across all replications, the exact reference model converged successfully with low standard deviations for the log-determinant, quadratic form, and predictive RMSE (Table~\ref{tab:vecchia_rep_summary}), establishing a stable baseline for comparison.The results for the decomposed approximation, summarized in Table~\ref{tab:vecchia_rep_summary}, reveal several key insights. First, for both Nearest-Neighbors and correlation-based (Corr) neighbour selection, the relative errors in the reconstructed inverse action $\bm{K}^{-1}y$, quadratic form $\bm{y}^\top \bm{K}^{-1}\bm{y}$, and log-determinant $\log|K|$ decrease monotonically as the neighbor size $m$ increases. This consistent decline, paired with small standard deviations across trials, provides empirical evidence that approximation errors do not propagate destructively through the model components. Second, the choice of ordering strategy significantly impacts the rate of convergence. Corr ordering substantially outperforms Nearest-Neighbor neighbour selection; for instance, at $m=20$, the mean relative error in the log-determinant under Corr conditioning ($3.2 \times 10^{-2}$) is nearly an order of magnitude lower than under Nearest-Neighbor ($2.2 \times 10^{-1}$). For $m \geq 30$, Corr conditioning achieves errors below $10^{-2}$, effectively recovering the exact likelihood at a fraction of the computational cost. Finally, these improvements in likelihood reconstruction translate directly to higher predictive accuracy. The mean RMSE on held-out high-fidelity observations decreases consistently as $m$ increases, with the lowest errors achieved under Corr conditioning. Collectively, these findings confirm that the proposed decomposed Vecchia strategy is numerically stable, statistically sound, and robust against random initializations.

\begin{table}[t]
\centering
\caption{Replicated validation of the decomposed Vecchia approximation. Results are averaged over 20 replications. Average exact RMSE is 0.688. Note this comparison does involve GLS adjustment.}
\label{tab:vecchia_rep_summary}
\resizebox{\linewidth}{!}{
\begin{tabular}{c c c c c c c}
\hline
Neighbour Selection & $m$ & $n_{\text{rep}}$ &
Mean rel.  $\|\bm{K}^{-1}\bm{y}\|$  (SD)&
Mean rel. $\log|\bm{K}|$ (SD) &
Mean rel. $\bm{y}^\top \bm{K}^{-1}\bm{y}$ (SD) &
Mean RMSE \\
\hline
Nearest-Neighbor & 10 & 20 & 0.464  (0.045) & 0.789 (0.144) & 0.0378 (0.0202) & 1.130 \\
Nearest-Neighbor & 20 & 20 & 0.345  (0.026) & 0.223 (0.045) & 0.0326 (0.0226) & 1.033 \\
Nearest-Neighbor & 30 & 20 & 0.315  (0.025) & 0.189 (0.040) & 0.0338 (0.0222) & 0.985 \\
Nearest-Neighbor & 40 & 20 & 0.298  (0.022) & 0.175 (0.037) & 0.0275 (0.0202) & 0.990 \\
Nearest-Neighbor & 60 & 20 & 0.208  (0.021) & 0.066 (0.011) & 0.0193 (0.0166) & 0.888 \\
\hline
Corr & 10 & 20 & 0.345  (0.026) & 0.247  (0.050) & 0.0221 (0.015)  & 0.957 \\
Corr & 20 & 20 & 0.220  (0.017) & 0.032  (0.007) & 0.0269 (0.021)  & 0.883 \\
Corr & 30 & 20 & 0.172  (0.019) & 0.0084 (0.006) & 0.0152 (0.010) & 0.790 \\
Corr & 40 & 20 & 0.124  (0.015) & 0.0055 (0.004) & 0.0131 (0.010) & 0.767 \\
Corr & 60 & 20 & 0.0748 (0.010) & 0.0074 (0.003) & 0.0088 (0.006) & 0.720 \\
\hline
\end{tabular}}
\end{table}

\begin{figure}
    \centering
    \includegraphics[width=0.9\linewidth]{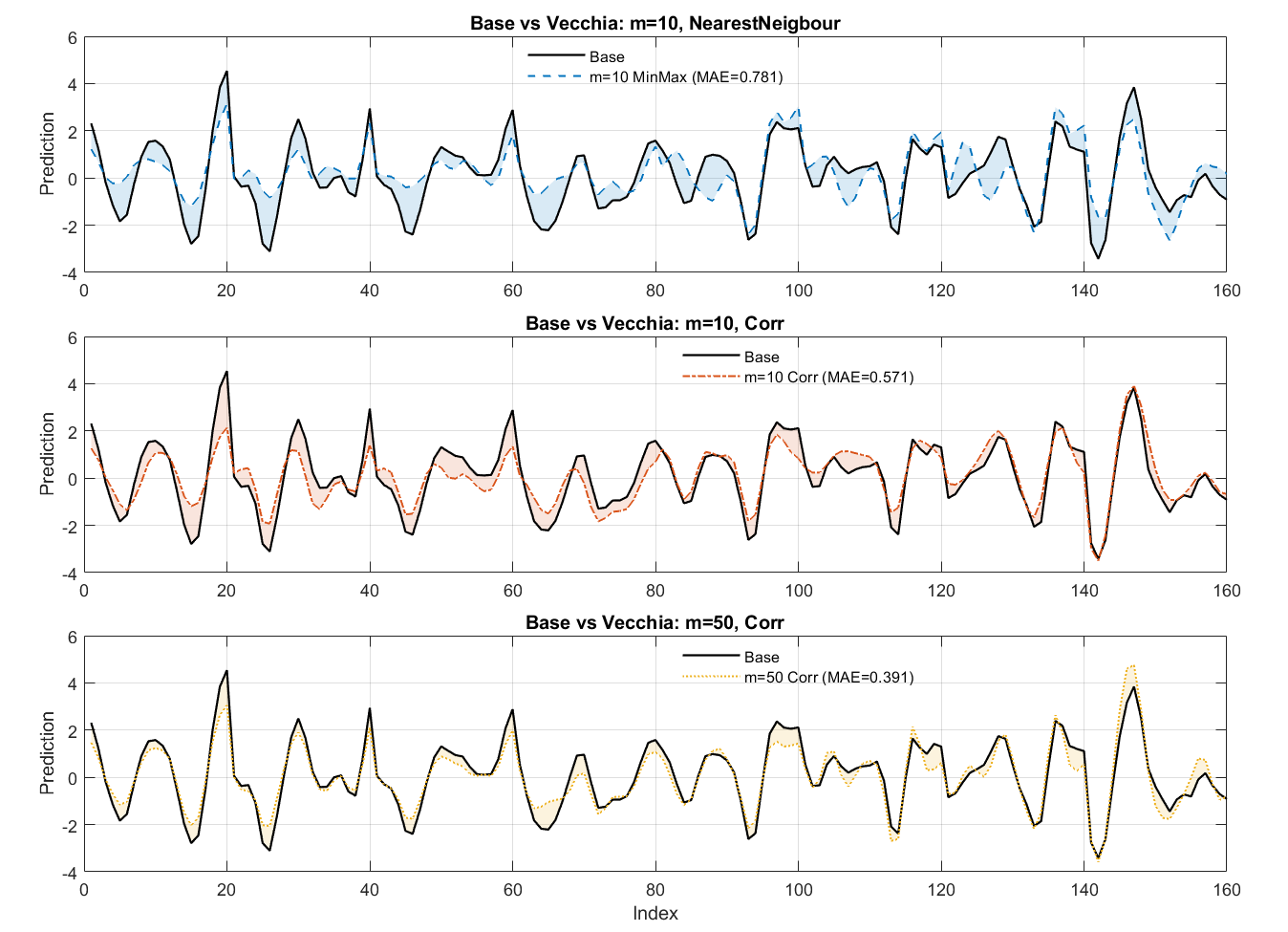}
    \caption{The figure illustrates predictions from a representative single replication of the experiment summarised in Table~\ref{tab:vecchia_rep_summary}. Predictions from the exact model are compared with those obtained using the decomposed Vecchia approximation under different neighbour sizes and conditioning strategies.}
    \label{fig:predictionVecchiaVSClassic}
\end{figure}

\noindent
\hangindent=-5mm
Examples of the plotted predictions of a single run of such an experiment are depicted in Figure \ref{fig:predictionVecchiaVSClassic}. Notice how generally the correlation conditioning is more precise than the Nearest-Neighbor.

 \subsection{Synthetic data experiment} \label{Synthetic_data_exp}
The dataset is designed to replicate the behavior of a continuous environmental variable. We utilize this setting to demonstrate that data-fusion methodologies are effective tools for reconstructing a target variable by integrating spatio-temporal information with multi-fidelity signals. The dataset size is intentionally constrained to facilitate exact GP inference without the need for sparse approximations. This allows the analysis to focus on whether our approximated framework achieves sufficient precision to offer a valuable alternative to standard MFGP models under realistic conditions, rather than assessing computational scalability. Details regarding the specific data generation process are provided in Appendix \ref{sec:sim_data}.

\noindent
\hangindent=-5mm
Performance is evaluated using a synthetic multi-fidelity spatio-temporal dataset generated through a controlled dynamic simulation. The experimental setup is defined on a regular spatial grid of size $6 \times 6$ ($n_{space} = 36$) over $10$ discrete time steps ($n_{time} = 10$), resulting in a full spatio-temporal domain of $n = 360$ observations. The simulation follows a linear auto-regressive multi-fidelity framework where the high-fidelity  process, $f_H$, is constructed as:
\begin{equation}
    f_H(\bm{s}, t) = \rho f_L(\bm{s}, t) + \delta(\bm{s}, t) + \epsilon_H,
\end{equation}
where $f_L$ is the low-fidelity (LF) latent process, $\rho = 0.60$ is the scaling coefficient, and $\delta$ represents the discrepancy process. Both $f_L$ and $\delta$ are modeled as Gaussian Processes with zero mean and separable  RBF kernels. The length scales $\ell_s$ and $\ell_t$ are analytically derived to satisfy target correlations of $0.72$ between adjacent spatial units and $0.80$ between successive time steps. The discrepancy variance is tested at levels of $\sigma_{\delta}^2 \in \{2, 4\}$.

\noindent
\hangindent=-5mm
To assess the models' capability for spatial extrapolation, the data is partitioned by station. Specifically, a training fraction of $0.3$ is applied to the spatial locations; thus, 12 stations (including their complete time series) are used for model fitting, while the remaining 24 stations are reserved entirely for out-of-sample testing. This setup ensures that predictive performance is measured on geographical locations unseen during the training phase.
 Each combination is replicated 100 times, with training and test spatial locations randomly re-drawn at each iteration to ensure robustness. We benchmark the proposed multi-fidelity approach against all possible GP model configurations regarding input features:
\begin{itemize}
    \item \textbf{GP-L:} $y_H \sim \mathrm{GP}(y_L(\bm{x}_H))$, where the model smooths the HF response using only the LF signal as an input ($y_L \rightarrow f \rightarrow y_H$). This is a 1D GP regression in LF-value space without using spatial coordinates. It isolates the value-to-value mapping $y_L$ to $y_H$ from space–time dependence. 
    \item \textbf{GP-3D:} $y_H \sim \mathrm{GP}(\bm{x}_H)$, where the model relies solely on the spatio-temporal coordinates ($[\bm{s}, t] \rightarrow f \rightarrow y_H$).
    \item \textbf{GP-4D:} $y_H \sim \mathrm{GP}(y_L(\bm{x}_H), \bm{x}_H)$, which jointly exploits the LF information and the spatio-temporal coordinates ($[y_L, \bm{s}, t] \rightarrow f \rightarrow y_H$). This tests whether adding $y_L$ as an extra covariate to a standard spatio-temporal improve the overall accuracy.
\end{itemize}
These variants cover the exhaustive set of alternatives for utilizing HF and LF data within a GP framework, allowing us to isolate the predictive contribution of physical coordinates versus cross-fidelity correlations. 

\subsubsection{Results of the synthetic data experiment} \label{result_synthetic}

Table \ref{tab:vecchia_performance_updated} reports the predictive performance under two noise levels, $\sigma_d^2 = 2$ and $\sigma_d^2 = 4$. Results are presented in terms of MAE, RMSE, and empirical coverage of the 95\% prediction intervals (COV95).

\noindent
\hangindent=-5mm
Across both noise settings, the \textit{Classic} (MFGP without approximation) approach consistently achieves the lowest MAE and RMSE, indicating superior point prediction accuracy. The performance gap is substantial compared to GP-3D, GP-L, and GP-4D, whose errors increase markedly as the noise variance grows. \textit{Vecchia\_40} represents the second-best performing method in terms of MAE and RMSE, maintaining relatively stable performance under increased noise. Indicating that our framework can provide reasonable approximation of the full model.

\noindent
\hangindent=-5mm
As expected, increasing the noise level from $\sigma_d^2 = 2$ to $\sigma_d^2 = 4$ leads to a deterioration in predictive accuracy for all methods. However, the degradation is more pronounced for GPs, while the Classic and Vecchia\_40 approaches show greater robustness.

\noindent
\hangindent=-5mm
Regarding uncertainty quantification, the empirical coverage (COV95) reveals a different pattern. GPs systematically have coverage probabilities well below the nominal 0.95 level, and the issue becomes more severe at higher noise levels. In contrast, both Classic and Vecchia\_40 provide coverage close to the nominal level. Notably, Vecchia\_40 achieves the most accurate coverage in both scenarios, slightly exceeding the nominal level, suggesting better-calibrated predictive intervals.

\noindent
\hangindent=-5mm
Overall, the results indicate that while the Classic method delivers the best point prediction accuracy, Vecchia\_40 provides a favorable trade-off between accuracy and uncertainty calibration. The remaining GP variants exhibit weaker robustness to noise and substantial under-coverage, highlighting limitations in their predictive uncertainty estimation under the considered experimental setting.

\noindent
\hangindent=-5mm
\begin{table}[h]
    \centering
    \scriptsize 
    \renewcommand{\arraystretch}{1.2} 
    \setlength{\tabcolsep}{5pt} 
    \rowcolors{2}{gray!10}{white} 
    \begin{tabular}{l|c|c|c|c|c|c}
        \toprule
        \textbf{Noise Level} & \textbf{Metric (sd)} & \textbf{GP-3D} & \textbf{GP-L} & \textbf{GP-4D} & \textbf{Classic} & \textbf{Vecchia\_v4} \\
        \midrule
        $\sigma_{d}^2=2$ 
        & MAE  
        & 2.143 (0.265) 
        & 2.346 (0.334) 
        & 2.257 (0.309) 
        & \cellcolor{green!40}1.343 (0.131) 
        & \cellcolor{yellow!40}1.527 (0.183) \\
        
        & RMSE 
        & 2.681 (0.320) 
        & 2.935 (0.399) 
        & 2.831 (0.370) 
        & \cellcolor{green!40}1.699 (0.170) 
        & \cellcolor{yellow!40}1.923 (0.224) \\
        
        & COV95 
        & 0.71 (0.058)     
        & 0.604 (0.069)     
        & 0.645 (0.057)       
        & 0.939 (0.033) 
        & \cellcolor{green!40}0.953 (0.039) \\
        
        \midrule
        
        $\sigma_{d}^2=4$ 
        & MAE  
        & 3.463 (0.461) 
        & 4.009 (0.547) 
        & 3.926 (0.523) 
        & \cellcolor{green!40}2.265 (0.314) 
        & \cellcolor{yellow!40}2.661 (0.463) \\
        
        & RMSE 
        & 4.324 (0.554) 
        & 5.023 (0.655) 
        & 4.919 (0.626) 
        & \cellcolor{green!40}2.885 (0.398) 
        & \cellcolor{yellow!40}3.345 (0.575) \\
        
        & COV95 
        & 0.61 (0.06)
        & 0.46 (0.06)
        & 0.49  (0.05)
        & 0.933 (0.048) 
        & \cellcolor{green!40}0.959 (0.055) \\
        
        \bottomrule
    \end{tabular}
    \caption{Performance comparison for the simulation study. Values are reported as mean (standard deviation). The neighborhood size is set to 40, observations are ordered temporally, and neighbor selection is based on correlation conditioning.}
    \label{tab:vecchia_performance_updated}
\end{table}

\subsection{Real data experiment} \label{Real_data_experiment}
We now turn to a real-world dataset to demonstrate the effectiveness of our methodology. The South Lombardy Wind Speed Dataset comprises wind speed measurements from the southern part of the Lombardy region in Italy. This dataset is particularly well-suited for evaluating modelling techniques, as it presents several meaningful challenges.

\noindent
\hangindent=-5mm
First, multi-fidelity methods without approximation are computationally infeasible on this dataset: the matrix inversion required for likelihood computations exceeds memory capacity in standard software. Second, the data concern wind speed, a variable commonly used in environmental modelling. Wind speed distributions are typically skewed and play a fundamental role in pollution dispersion modelling, making this dataset representative of broader environmental use cases. Third, the region exhibits significant variation in station density across a largely flat terrain with minimal natural obstructions, but a non-uniform spatial correlation decay. These characteristics make it an ideal testbed for assessing wind dynamics in a context that is both structured and complex.

\noindent
\hangindent=-5mm
The dataset consists of two primary data sources:  
\textit{ERA5 Global Reanalysis Data} \citep{ERA5}, providing LF hourly wind speed estimates on a gridded domain for January 2022. This source offers a broad-scale representation of regional wind trends.  
\textit{ARPA Lombardia Monitoring Network} \citep{ARPA_Lombardia, earth3010013}, supplying HF wind speed observations from 18 ground-based stations located in southern Lombardy. These measurements are highly accurate and essential for capturing local-scale variability. The HF data were obtained using the R package ARPALData \citep{maranzano2024arpaldata}.

\noindent
\hangindent=-5mm
To align the two data sources spatially, we employed a nearest-neighbour matching approach: each ARPA monitoring station was paired with its closest ERA5 grid cell. The resulting merged dataset contains 26,784 observations, forming a multi-fidelity covariance matrix with over 717 million entries.

\begin{figure}
    \centering
    \includegraphics[width=0.9\linewidth]{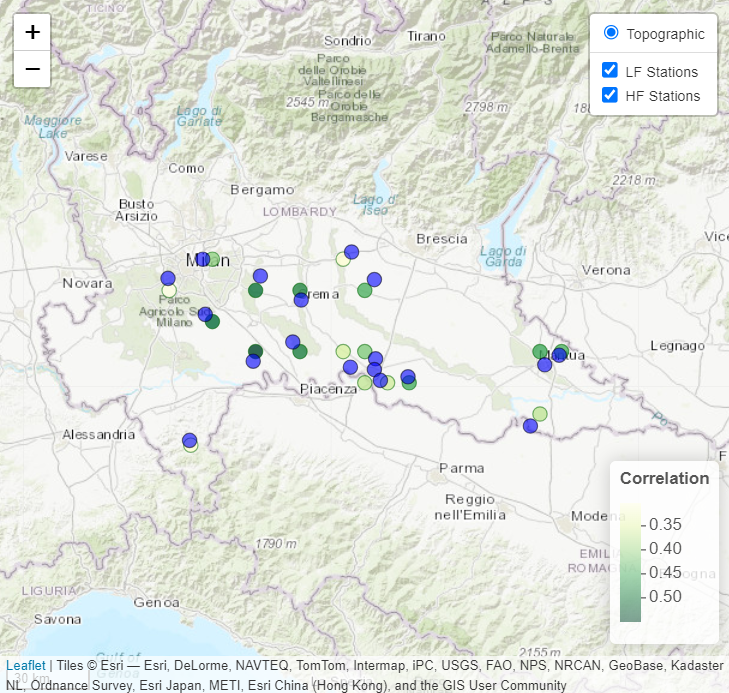}
    \caption{Illustration of the South Lombardy dataset. The blue dots depicts the position of ARPA monitoring station, in light green instead the position of ERA reanalysis data grid cell center. The legend reports the correlation coeefficient between the two datasources.}
    \label{fig:lombardy}
\end{figure}

\noindent
\hangindent=-5mm
Figure~\ref{fig:lombardy} shows the spatial locations of the monitoring stations alongside their corresponding ERA5 grid cells. The linear correlation coefficients between the two data sources are also reported, providing an initial measure of cross-fidelity consistency.
 
\subsubsection{Design of real data experiment}

We compare a single-fidelity spatio-temporal Gaussian process model with a  set of multi-fidelity Gaussian process  models that differ in the treatment of the cross-fidelity integration function $\rho(\bm{s})$ and GLS estimation. The models considered in this experiment are summarised in Table~\ref{tab:model_type}.

As a single-fidelity benchmark, we consider an approximate spatio-temporal Gaussian process regression applied to the three spatio-temporal coordinates, hereafter 
denoted as \textbf{GP-3D (approx)}. This model relies exclusively on 
HF data and does not exploit information from the low-fidelity (LF) source. We choosed the \textbf{GP-3D} as benchmark for two reasons. First it was the best-performing model in the synthetic experiments 
(see Table~\ref{tab:vecchia_performance_updated}), second preliminary analyses whose results are stored in the supplementary material conducted on smaller subsets of the real dataset indicated that it provides the most robust  baseline among the considered GP alternatives.

To assess the relative impact of mean-structure flexibility, cross-fidelity scaling non-stationarity, and latent space transformations, we evaluate the multi-fidelity Gaussian process (MFGP) framework across six distinct configurations. These variants, summarized in Table \ref{tab:model_type}, are defined by the intersection of three methodological axes:

\begin{enumerate}
    \item \textbf{Mean Structure Strategy:} We compare two implementations of the Generalised Least Squares (GLS) procedure. The \textit{global GLS} approach assumes a spatially homogeneous baseline shift, utilizing global intercepts $\beta_L$ and $\beta_H$. The \textit{Adaptive GLS} variant incorporates a spatial linear trend by augmenting the design matrix $\bm{G}$ with geographic coordinates. This allows the intercepts to vary as a first-order polynomial of space: $\beta_f(\bm{s}) = \beta_{f,0} + \beta_{f,lat}s_{lat} + \beta_{f,lon}s_{lon}$, effectively centering the process residuals against regional systematic biases.
    
    \item \textbf{Cross-Fidelity Scaling ($\rho$):} We distinguish between a \textit{Constant} $\rho$, implying a uniform rescaling across the domain, and a \textit{GP-based Adaptive} $\rho(\bm{s})$. In the latter, local empirical estimates of temporal co-variability are regularized through a secondary Gaussian process regression to ensure spatial smoothness and robust information transfer between fidelities.
    
    \item \textbf{Warping:} Following the methodology of \citet{colombo2025warped}, certain configurations employ a \textit{Warped} latent space. This monotone transformation maps skewed observations into a latent Gaussian space for inference, ensuring that back-transformed predictive means remain strictly positive and physically consistent.
\end{enumerate}
In the implementation of MFGP models variant we decided to focus on those showed an high degree of differentiation.

Model performance is evaluated using a leave-one-station-out cross-validation 
(LOSO-CV) strategy, following the approach of \citet{otto2024spatiotemporal}. 
For each station $i$, model parameters are estimated using the training set
\[
\mathcal{D}_{\text{train}}^{(i)} = \mathcal{D}_L \cup \mathcal{D}_H 
\setminus \left( x_H^{(i)}, y_H^{(i)} \right),
\]
where $\mathcal{D}_L$ denotes the complete LF dataset, $\mathcal{D}_H$ the complete 
HF dataset, and $\left( x_H^{(i)}, y_H^{(i)} \right)$ corresponds to the HF 
observations from the held-out station. Each trained model is then evaluated on the 
excluded station to assess predictive accuracy and generalization performance. As said in section \ref{Real_data_experiment} all the HF and LF station are matched with nearest neighbour matching.

\begin{table}[h]
    \centering
    \scriptsize
    \renewcommand{\arraystretch}{1.2}
    \setlength{\tabcolsep}{6pt}
    \rowcolors{2}{gray!10}{white}
    \begin{tabular}{l|c|c|c}
        \toprule
        \textbf{Model ID} & \textbf{Mean Structure} & $\rho(\bm{s})$  & Input Warping  \\
        \midrule
        GP-3D (Approx) & NA   & NA                             & NO  \\
        $MFGP_{gc}$& Global GLS           & Constant            & NO    \\
        $MFGP_{ac}$& Adaptive GLS        & Constant            & NO   \\
        $MFGP_{gWc}$& Global GLS          & Constant            & YES  \\
        $MFGP_{aWc}$& Adaptive GLS       & Constant            & YES  \\
        $MFGP_{gGP}$      & Global GLS    & GP-based (empirical)& NO  \\
        $MFGP_{aGP}$      & Adaptive GLS & GP-based (empirical)& NO  \\
        \bottomrule
    \end{tabular}
    \caption{Experimental configurations. Models are distinguished by their mean structure (fixed vs. adaptive GLS offsets), the specification of the cross-fidelity parameter $\rho(\bm{s})$, and whether input warping is applied.}
    \label{tab:model_type}
\end{table}

\subsubsection{Results of real data experiments}

The performance of the six MFGP variants was evaluated across 18 stations using five key metrics: Mean Absolute Error (MAE), Root Mean Square Error (RMSE), Pearson Correlation (Corr), 95\% Prediction Interval Coverage Probability ($\text{PICP}_{95}$), and Negative Log-Marginal Likelihood (NLML). Table \ref{tab:summary_results} summarizes the aggregated performance for each configuration.

\begin{table}[h]
    \centering
    \scriptsize
    \renewcommand{\arraystretch}{1.2}
    \setlength{\tabcolsep}{5pt}
    \rowcolors{2}{gray!10}{white}
    \begin{tabular}{lcccccc}
        \toprule
        \textbf{Model ID} & \textbf{Count} & \textbf{MAE} & \textbf{RMSE} & \textbf{Corr} & \textbf{PICP$_{95}$} & \textbf{NLML} \\
        \midrule
        $GP-3D$   & 18 &0.3974 & 0.5245 &0.77 & 93.84\% \\
        $MFGP_{gc}$   & 18 & \textbf{0.1906} & \textbf{0.2429} & \textbf{0.9351} & 89.06 & 2790.6 \\
        $MFGP_{ac} $  & 18 & 0.1935 & 0.2477 & 0.9312 & 58.28 & 2738.5 \\
        $MFGP_{gWc}$ & 18 & 0.1949 & 0.2487 & 0.9307 & \textbf{91.39} & 3323.3 \\
        $MFGP_{aWc}$  & 18 & 0.2005 & 0.2552 & 0.9304 & 76.94 & 3274.5 \\
        $MFGP_{gGP}$    & 18 & 0.3203 & 0.4007 & 0.8729 & 83.78 & 2729.3 \\
        $MFGP_{aGP}$   & 18 & 0.4446 & 0.5598 & 0.8854 & 83.17 & 2786.3 \\
        \bottomrule
    \end{tabular}
    \caption{Aggregated performance metrics across 18 validation stations. Bold values indicate the best performance in each category.}
    \label{tab:summary_results}
\end{table}

The experimental results reveal a distinct trade-off between deterministic point accuracy and statistical reliability. While the $MFGP_{gc}$ model achieved the lowest overall error (MAE: 0.1906), the integration of more complex mean structures and latent space transformations yielded significant insights into model stability.

\paragraph{Calibration vs. Accuracy:} 
A critical observation arises from the $MFGP_{ac}$ variant. Despite maintaining a competitive MAE (0.1935), its uncertainty calibration collapsed to a PICP$_{95}$ of 58.28\%. This suggests that incorporating a spatial trend in the GLS mean without further regularization causes the model to become overconfident, attributing too much variance to the deterministic component and artificially shrinking the predictive intervals. Conversely, the warped variant $MFGP_{cWc}$ provided the most calibrated intervals (91.39\%), demonstrating the role of the Colombo et al. (2025) transformation in stabilizing inference for skewed wind-speed data.

\begin{figure}
    \centering
    \includegraphics[width=1.1\linewidth]{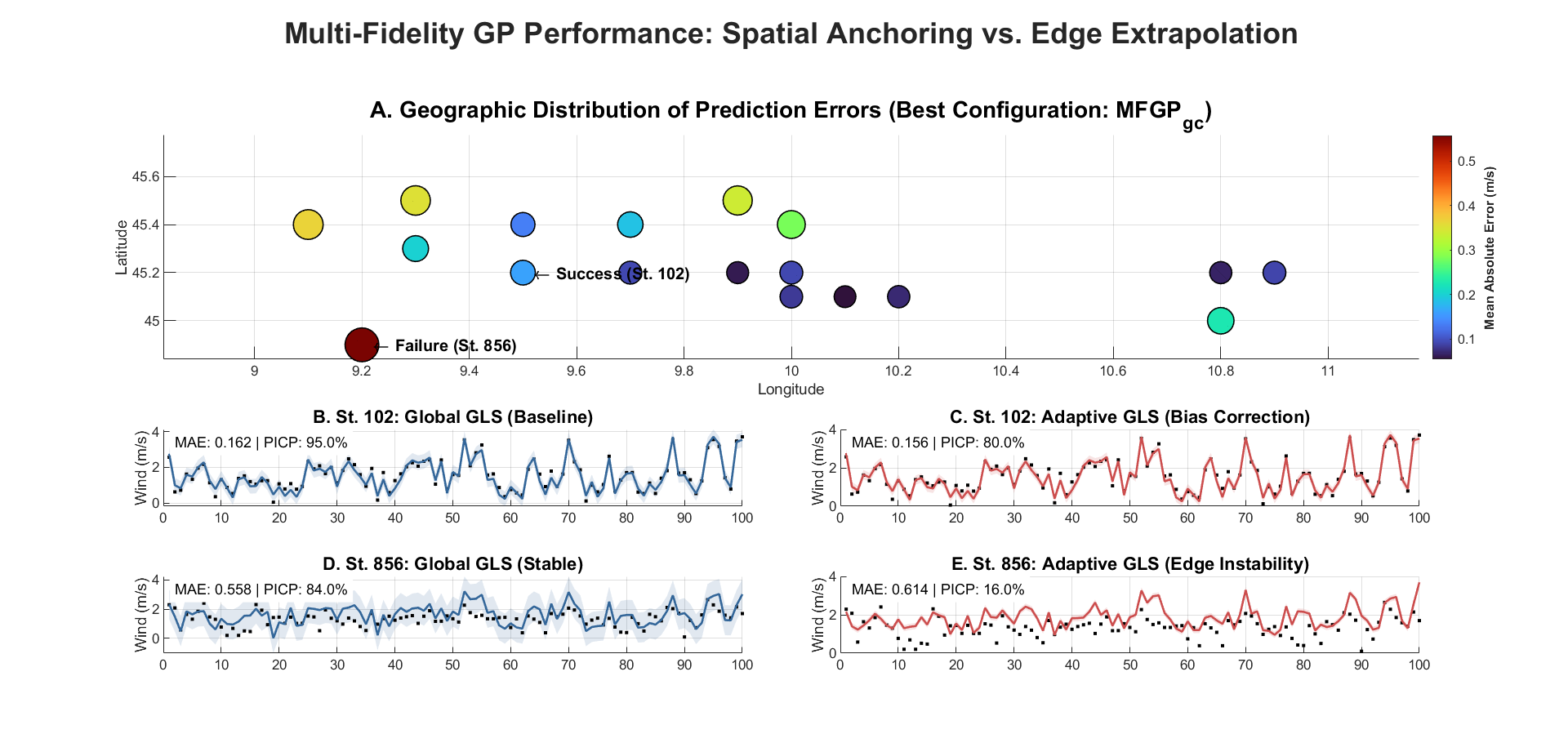}
    \caption{Spatial and temporal performance analysis of the MFGP framework across the Lombardy sensor network. (Top) Geographic distribution of the Mean Absolute Error (MAE) for the best-performing configuration ($MFGP_{gc}$). The marker size and color intensity represent the error magnitude, highlighting a trend of increased residuals at the domain boundaries (e.g., Station 856) compared to central clusters. (Bottom - Row 1) Comparative time-series for Station 102, illustrating a \enquote{Success Case} for the Adaptive GLS approach; the spatial trend effectively corrects local sensor bias, centering the prediction on the observations. (Bottom - Row 2) Comparative time-series for Station 856, illustrating \enquote{Edge Instability}; the linear spatial trend, lacking geographic anchoring, over-extrapolates the mean structure, leading to higher residuals and narrower, overconfident prediction intervals (reduced PICP) compared to the more robust Global GLS baseline.}
    \label{fig:comparison_MFGP}
\end{figure}

\paragraph{Impact of Adaptive GLS:}
The station-by-station analysis indicates that the Adaptive GLS (spatial trend) approach is highly sensitive to the spatial distribution of sensors. In stations characterized by high local bias between low- and high-fidelity signals, the adaptive variants often outperformed the global intercept models. However, in regions with sparse observations, the linear spatial trend risked over-extrapolation, which explains the slightly higher average RMSE compared to Global GLS counterparts. These elements are evidenced in Figure \ref{fig:comparison_MFGP}, where we reported the comparison of the prediction of an adaptive GLS versus a constant adjustment for a station in the centre and one in the neighbour.

\begin{figure}
    \centering
    \includegraphics[width=1\linewidth]{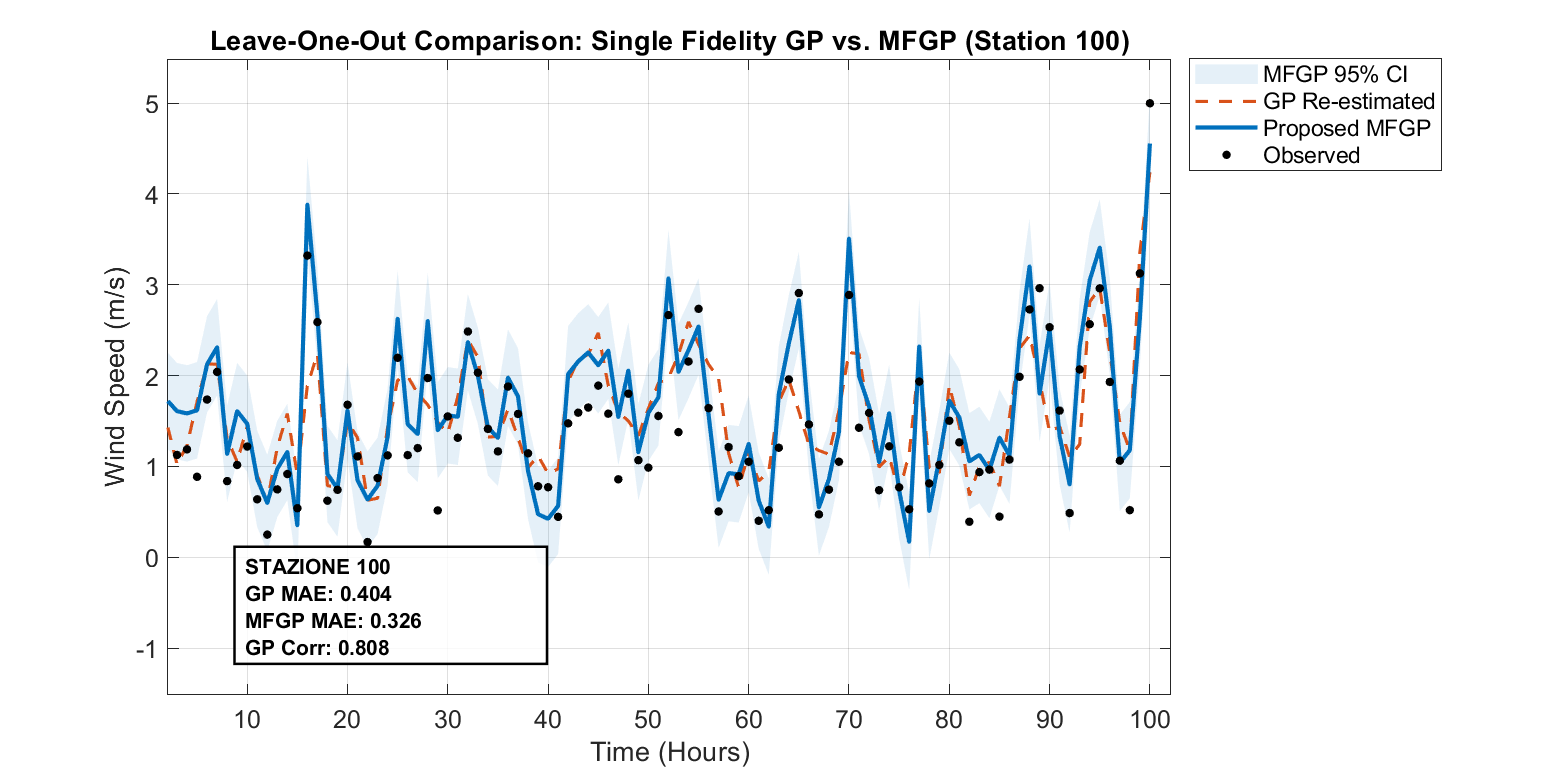}
    \caption{The figure depicts the comparison between $MFGP_{gc}$ and the GP-3D (without approximation) in red on test station 100.}
    \label{fig:GP_vs_MFGP}
\end{figure}

\paragraph{Gaussian process prediction:}
The oversmoothing in our Gaussian process predictions stems largely from the approximation techniques used to handle the dataset size. However, the MFGP framework consistently outperforms standard GPs, even when the latter are fitted without approximations on smaller data subsets. An example of a comparison of an not approximate GP with an approximated $MFGP_{gc}$ is depicted in Figure \ref{fig:GP_vs_MFGP}.

\paragraph{Non-Stationary Scaling ($\rho$):}
The configurations utilizing GP-based adaptive scaling exhibited significantly higher errors. This degradation in performance suggests that the local temporal co-variability used to estimate $\rho(\bm{s})$ may be susceptible to noise. The global constant scaling factor appears more robust for this specific dataset, likely because the relationship between the two fidelities remains relatively stable across the sampled geographic domain.

\section{Conclusions}
This paper developed a scalable multi-fidelity Gaussian process framework for spatio-temporal environmental data fusion, motivated by a common practical setting: low-fidelity products (such as reanalysis or satellite estimates) are abundant but noisy, while high-fidelity in-situ measurements are accurate but sparse and geographically uneven. The central goal was to obtain high-resolution reconstructions that preserve local variability and realistic uncertainty, without the computational barriers that typically prevent multi-fidelity Gaussian processes from being trained with likelihood-based inference on large datasets.

The key methodological outcome is a practical way to make multi-fidelity Gaussian processes compatible with Vecchia approximations while retaining full marginal-likelihood evaluation. Instead of approximating the full multi-fidelity covariance directly—which is challenging due to cross-fidelity dependence—the approach decomposes the model into a low-fidelity latent process and an independent discrepancy process. Vecchia conditioning is then applied to these latent components rather than to the full joint model, and the full multi-fidelity likelihood is computed efficiently through a reconstruction strategy that avoids forming the dense covariance matrix. This makes the framework computationally feasible in large spatio-temporal problems while also improving numerical stability.

A second methodological contribution concerns the mean structure. The paper shows that multi-fidelity models can otherwise absorb systematic baseline differences between low- and high-fidelity sources into the inferred cross-fidelity dependence, which can distort the interpretation and degrade prediction. To prevent this, the framework removes fidelity-specific offsets using a generalized least squares strategy before likelihood evaluation and prediction. This centering step encourages the covariance structure to represent genuine spatio-temporal dependence and cross-fidelity information transfer rather than persistent level shifts.

The framework was evaluated through three experiments designed to test stability, accuracy, and practical value. The first experiment focused on validating the decomposed Vecchia strategy itself. Across repeated replications, approximation errors in likelihood components and predictive accuracy improved consistently as the neighbor size increased, indicating that the decomposition and reconstruction do not introduce unstable behavior. The experiment also showed that conditioning choices matter: correlation-based neighbor selection yielded substantially faster convergence toward the exact likelihood than a simple geometric approach, and it translated into better predictive performance at comparable approximation levels.

The synthetic-data experiment demonstrated that the proposed approximated multi-fidelity model remains a faithful substitute for exact multi-fidelity inference under controlled conditions. The Vecchia-based multi-fidelity model closely tracked the classic (non-approximated) multi-fidelity Gaussian process, with only small losses in error metrics, while consistently outperforming all single-fidelity Gaussian process baselines. This supports the main claim that the approximation preserves the essential advantage of multi-fidelity modeling—namely, leveraging both spatio-temporal structure and cross-fidelity dependence—while enabling scalability.

The real-data experiment on wind speed in southern Lombardy provided the strongest evidence of practical impact. On a dataset size for which exact multi-fidelity inference is computationally infeasible, the proposed framework delivered large improvements over the single-fidelity spatio-temporal Gaussian process benchmark. Across leave-one-station-out validation, multi-fidelity models achieved markedly lower errors and substantially higher correlations, indicating better reconstruction of local temporal patterns and spatial variability at unseen stations. The comparison among multi-fidelity variants revealed important practical trade-offs. The simplest multi-fidelity configuration—with fixed fidelity offsets and constant cross-fidelity scaling—was the most reliable overall in point prediction accuracy and correlation for this dataset. Introducing a more flexible spatial mean structure sometimes improved performance at individual stations, but it could also lead to instability and overconfidence in poorly supported regions, reflected by a sharp drop in prediction interval coverage in the aggregated results. Warping improved calibration and yielded the best uncertainty coverage among the multi-fidelity variants, confirming that distributional skewness in wind speed can meaningfully affect uncertainty quantification. Finally, the spatially varying cross-fidelity scaling strategy underperformed on this application, suggesting that the empirical estimation of spatially varying coupling can be sensitive to noise and may require stronger regularization or alternative constructions to be consistently beneficial.

Overall, the paper demonstrates that multi-fidelity spatio-temporal Gaussian processes can be made scalable, stable, and fully likelihood-based by combining latent-process decomposition with Vecchia approximations and efficient likelihood reconstruction. The experiments show that the resulting method reproduces exact multi-fidelity behavior in controlled settings and delivers substantial predictive gains in a realistic environmental application where large data size and uneven sampling make standard approaches inadequate. The results also highlight that, in real monitoring networks, robustness often comes from disciplined model structure: controlling mean offsets and avoiding overly flexible components without sufficient support can be as important as improving the covariance model itself. This framework therefore provides a practical foundation for large-scale environmental reconstruction tasks and a flexible basis for future extensions aimed at more robust non-stationary cross-fidelity coupling and improved uncertainty calibration under heterogeneous sensor coverage.

\section*{Acknowledgements}

This article is distributed under the terms of the Creative Commons Attribution License (CC BY), which permits unrestricted use, distribution, and reproduction in any medium, provided the original author and source are credited. For more information, visit 
 \url{https://creativecommons.org/licenses/by/4.0/}.

\bigskip
\begin{center}
{\large\bf SUPPLEMENTARY MATERIAL}
\end{center}

\begin{description}
\item[Title:] Pietrostat193/
Public-Vecchia-Approximation-for-multifidelity-models.
\item \href{https://github.com/Pietrostat193/Public-Vecchia-Approximation-for-multifidelity-models}{\texttt{GitHub Repo Link}}
\item[Matlab-code:] The repository containing code to test the method.
\item[Datasets:] All the datasets used in this article: South Lombardy, Whole Lombardy, Synthetic data generation function, plus additional toy dataset used for illustrative purposes.
\end{description}

\bibliographystyle{chicago}

\bibliography{Bibliography-MM-MC}

@article{perdikaris2017nonlinear,
  title={Nonlinear information fusion algorithms for data-efficient multi-fidelity modelling},
  author={Perdikaris, Paris and Raissi, Maziar and Damianou, Andreas and Lawrence, Neil D and Karniadakis, George Em},
  journal={Proceedings of the Royal Society A: Mathematical, Physical and Engineering Sciences},
  volume={473},
  number={2198},
  pages={20160751},
  year={2017},
  publisher={The Royal Society Publishing}
}

@article{raissi2016deep,
  title={Deep multi-fidelity Gaussian processes},
  author={Raissi, Maziar and Karniadakis, George},
  journal={arXiv preprint arXiv:1604.07484},
  year={2016}
}

@article{sauer2023vecchia,
  title={Vecchia-approximated deep Gaussian processes for computer experiments},
  author={Sauer, Annie and Cooper, Andrew and Gramacy, Robert B},
  journal={Journal of Computational and Graphical Statistics},
  volume={32},
  number={3},
  pages={824--837},
  year={2023},
  publisher={Taylor \& Francis}
}

@article{babaee2020multifidelity,
  title={A multifidelity framework and uncertainty quantification for sea surface temperature in the massachusetts and cape cod bays},
  author={Babaee, Hessam and Bastidas, C and Defilippo, Michael and Chryssostomidis, C and Karniadakis, GE},
  journal={Earth and Space Science},
  volume={7},
  number={2},
  pages={e2019EA000954},
  year={2020},
  publisher={Wiley Online Library}
}

@article{colombo2025warped,
  title={Warped multifidelity Gaussian processes for data fusion of skewed environmental data},
  author={Colombo, Pietro and Miller, Claire and Yang, Xiaochen and O’Donnell, Ruth and Maranzano, Paolo},
  journal={Journal of the Royal Statistical Society Series C: Applied Statistics},
  pages={qlaf003},
  year={2025},
  publisher={Oxford University Press UK}
}

@article{otto2024spatiotemporal,
  title={Spatiotemporal modelling of PM 2.5 concentrations in Lombardy (Italy): a comparative study},
  author={Otto, Philipp and Fusta Moro, Alessandro and Rodeschini, Jacopo and Shaboviq, Qendrim and Ignaccolo, Rosaria and Golini, Natalia and Cameletti, Michela and Maranzano, Paolo and Finazzi, Francesco and Fass{\`o}, Alessandro},
  journal={Environmental and Ecological Statistics},
  volume={31},
  number={2},
  pages={245--272},
  year={2024},
  publisher={Springer}
}

@misc{ARPA_Lombardia,
  author       = {{Agenzia Regionale per la Protezione dell'Ambiente Lombardia}},
  title        = {Dati e Indicatori - ARPA Lombardia},
  year         = {2025},
  url          = {https://www.arpalombardia.it/dati-e-indicatori/},
  note         = {Accessed: 2025-03-19}
}

@article{Kennedy2000,
  author    = {M. C. Kennedy and A. O'Hagan},
  title     = {Predicting the Output from a Complex Computer Code When Fast Approximations Are Available},
  journal   = {Biometrika},
  volume    = {87},
  number    = {1},
  pages     = {1--13},
  year      = {2000},
  month     = {March},
  publisher = {Oxford University Press},
  doi       = {10.1093/biomet/87.1.1}
}

@inproceedings{damianou2013deep,
  title={Deep gaussian processes},
  author={Damianou, Andreas and Lawrence, Neil D},
  booktitle={Artificial intelligence and statistics},
  pages={207--215},
  year={2013},
  organization={PMLR}
}

@article{katzfuss2021general,
  title={A general framework for Vecchia approximations of Gaussian processes},
  author={Katzfuss, Matthias and Guinness, Joseph},
  journal={Statistical Science},
  volume={36},
  number={1},
  pages={124--141},
  year={2021},
  publisher={JSTOR}
}

@misc{ERA5,
  author       = {{Copernicus Climate Change Service (C3S)}},
  title        = {ERA5: Fifth Generation of ECMWF Atmospheric Reanalyses of the Global Climate},
  year         = {2025},
  url          = {https://cds.climate.copernicus.eu/datasets/reanalysis-era5-single-levels?tab=overview},
  note         = {Accessed: 2025-03-19}
}

@article{toal2011efficient,
  title={Efficient multipoint aerodynamic design optimization via cokriging},
  author={Toal, David JJ and Keane, Andy J},
  journal={Journal of Aircraft},
  volume={48},
  number={5},
  pages={1685--1695},
  year={2011}
}

@article{peherstorfer2019multifidelity,
  title={Multifidelity Monte Carlo estimation with adaptive low-fidelity models},
  author={Peherstorfer, Benjamin},
  journal={SIAM/ASA Journal on Uncertainty Quantification},
  volume={7},
  number={2},
  pages={579--603},
  year={2019},
  publisher={SIAM}
}

@article{kaps2022hierarchical,
  title={A hierarchical kriging approach for multi-fidelity optimization of automotive crashworthiness problems},
  author={Kaps, Arne and Czech, Catharina and Duddeck, Fabian},
  journal={Structural and Multidisciplinary Optimization},
  volume={65},
  number={4},
  pages={114},
  year={2022},
  publisher={Springer}
}

@article{zhang2022multi,
  title={Multi-scale Vecchia approximations of Gaussian processes},
  author={Zhang, Jingjie and Katzfuss, Matthias},
  journal={Journal of Agricultural, Biological and Environmental Statistics},
  volume={27},
  number={3},
  pages={440--460},
  year={2022},
  publisher={Springer}
}

@article{kundig2024iterative,
  title={Iterative methods for vecchia-laplace approximations for latent gaussian process models},
  author={K{\"u}ndig, Pascal and Sigrist, Fabio},
  journal={Journal of the American Statistical Association},
  pages={1--14},
  year={2024},
  publisher={Taylor \& Francis}
}

@article{cheng2021hierarchical,
  title={Hierarchical Bayesian nearest neighbor co-kriging Gaussian process models; an application to intersatellite calibration},
  author={Cheng, Si and Konomi, Bledar A and Matthews, Jessica L and Karagiannis, Georgios and Kang, Emily L},
  journal={Spatial Statistics},
  volume={44},
  pages={100516},
  year={2021},
  publisher={Elsevier}
}

@article{rambelli2025accuracy,
  title={An accuracy-runtime trade-off comparison of scalable Gaussian process approximations for spatial data},
  author={Rambelli, Filippo and Sigrist, Fabio},
  journal={arXiv preprint arXiv:2501.11448},
  year={2025}
}

@article{cheng2024recursive,
  title={Recursive nearest neighbor co-kriging models for big multi-fidelity spatial data sets},
  author={Cheng, Si and Konomi, Bledar A and Karagiannis, Georgios and Kang, Emily L},
  journal={Environmetrics},
  volume={35},
  number={4},
  pages={e2844},
  year={2024},
  publisher={Wiley Online Library}
}

@article{le2014recursive,
  title={Recursive co-kriging model for design of computer experiments with multiple levels of fidelity},
  author={Le Gratiet, Loic and Garnier, Josselin},
  journal={International Journal for Uncertainty Quantification},
  volume={4},
  number={5},
  year={2014},
  publisher={Begel House Inc.}
}

@article{vecchia1988estimation,
  title={Estimation and model identification for continuous spatial processes},
  author={Vecchia, Aldo V},
  journal={Journal of the Royal Statistical Society Series B: Statistical Methodology},
  volume={50},
  number={2},
  pages={297--312},
  year={1988},
  publisher={Oxford University Press}
}

@article{datta2016hierarchical,
  title={Hierarchical nearest-neighbor Gaussian process models for large geostatistical datasets},
  author={Datta, Abhirup and Banerjee, Sudipto and Finley, Andrew O and Gelfand, Alan E},
  journal={Journal of the American Statistical Association},
  volume={111},
  number={514},
  pages={800--812},
  year={2016},
  publisher={Taylor \& Francis}
}

@article{lalchand2022sparse,
  title={Sparse gaussian process hyperparameters: Optimize or integrate?},
  author={Lalchand, Vidhi and Bruinsma, Wessel and Burt, David and Rasmussen, Carl Edward},
  journal={Advances in Neural Information Processing Systems},
  volume={35},
  pages={16612--16623},
  year={2022}
}

@article{maranzano2024arpaldata,
  title={ARPALData: an R package for retrieving and analyzing air quality and weather data from ARPA Lombardia (Italy)},
  author={Maranzano, Paolo and Algieri, Andrea},
  journal={Environmental and Ecological Statistics},
  volume={31},
  number={2},
  pages={187--218},
  year={2024},
  publisher={Springer}
}

@article{FernandezGodino2023,
  title   = {Review of multi-fidelity models},
  author  = {Fern{\'a}ndez-Godino, M. Giselle},
  journal = {Advances in Computational Science and Engineering},
  volume  = {1},
  number  = {4},
  pages   = {351--400},
  year    = {2023},
  doi     = {10.3934/acse.2023015},
  url     = {https://www.aimsciences.org/article/id/656460fbeca2737fc4ef5257}
}

@article{brevault2020overview,
  title={Overview of Gaussian process based multi-fidelity techniques with variable relationship between fidelities, application to aerospace systems},
  author={Brevault, Lo{\"\i}c and Balesdent, Mathieu and Hebbal, Ali},
  journal={Aerospace Science and Technology},
  volume={107},
  pages={106339},
  year={2020},
  publisher={Elsevier}
}

@article{earth3010013,
author = {Maranzano, Paolo},
title = {Air Quality in Lombardy, Italy: An Overview of the Environmental Monitoring System of ARPA Lombardia},
journal = {Earth},
volume = {3},
year = {2022},
number = {1},
pages = {172--203},
url = {https://www.mdpi.com/2673-4834/3/1/13},
issn = {2673-4834},
DOI = {10.3390/earth3010013}
}

\newpage 
\appendix 

\counterwithin{figure}{section}
\counterwithin{table}{section}
\counterwithin{equation}{section}

\paragraph{Methodological supplements}
\section{Understanding the effect of sparsity on positive semidefiniteness} \label{PositiveSemiDefinite}

The algorithm described in Section \ref{NewMethod} requires the evaluation of a joint covariance structure modified by the cross-fidelity scaling $\rho(\bm{s})$. Mathematically, this involves operations effectively equivalent to a transformation of the kernel matrix $k(\bm{x}, \bm{x}')$. In a dense implementation, such operations are prone to numerical instability, as small perturbations in $\rho(\bm{s})$ can lead to a loss of the positive-semidefinite (PSD) property in the joint system.

By contrast, the Vecchia-approximated framework we propose acts as a form of \textbf{structural regularization}. In this context, the sparsity of the precision matrix (and the underlying local structure of the covariance) provides several critical advantages:

\begin{itemize}
    \item \textbf{Spectrum Stabilization:} High-dimensional dense covariance matrices often suffer from a high condition number due to the accumulation of small, spurious long-range correlations. The Vecchia approximation implicitly regularizes the spectrum by enforcing a local Markov property, which zeros out these long-range entries and prevents the matrix from becoming nearly singular.
    
    \item \textbf{Local Dependency Preservation:} Sparsity in the covariance structure reflects the physical reality of local interactions in spatio-temporal data. By restricting the influence of $\rho(\bm{s})$ to local neighbor sets, we ensure that the resulting joint matrix preserves structured dependencies without propagating numerical errors across the entire domain.
    
    \item \textbf{Reduced Error Accumulation:} Operations on sparse factors (such as the $B$ and $D$ matrices in the Vecchia scheme) are significantly less likely to introduce catastrophic rounding errors during the Cholesky factorization. This is due to the reduced number of floating-point operations required compared to dense matrix-vector multiplications, which helps preserve the numerical integrity of the PSD constraint.
\end{itemize}

It is therefore unsurprising that our improved MFGP algorithm demonstrated superior convergence rates. By leveraging sparsity as an implicit regularizer, the model remains robust even in the presence of complex non-stationary mean structures and spatially-varying scaling functions, where standard dense implementations typically fail to converge.

\section{Effect of ordering and conditioning on the precision matrix $\mathbf{H}$}
\label{sec:supp_ordering_H}

The sparsity and apparent bandedness of the precision matrix
\[
\bm{H} = \boldsymbol{\Sigma}_w^{-1} + \bm{A}^\top \bm{D}_\epsilon^{-1} \bm{A},
\]
depend on the ordering used in the Vecchia approximation, since conditioning
sets are restricted to preceding indices.
In all experiments in the main paper, observations are ordered by spatial
location (space-major ordering), with time varying fastest within each
location, and correlation-based Vecchia conditioning with $m=15$ neighbors
is employed.
To assess the sensitivity of $\bm{H}$ to alter Nearest-Neighbors, we
recomputed $\bm{H}$ under several plausible schemes: space-major
(ordering used throughout the paper), time-major, space-major, and random
orderings.

Table~\ref{tab:H_ordering} reports sparsity and factorization statistics for
a representative simulated space--time data set with $n_L = n_H = 720$.
Across orderings, the number of nonzeros in $\bm{H}$ ranges from
$8.1\times 10^4$ to $9.6\times 10^4$ (densities $3.9\%$--$4.6\%$).
After applying a fill-reducing approximate minimum degree (AMD) permutation,
the number of nonzeros in the sparse Cholesky factor ranges from
$3.17\times 10^5$ to $4.26\times 10^5$, indicating moderate variation but
consistently sparse factors.
While time-major and random orderings lead to increased fill-in relative to
the space-major ordering, the overall sparsity remains of the same order
of magnitude.

\begin{figure}[t]
\centering
\includegraphics[width=\textwidth]{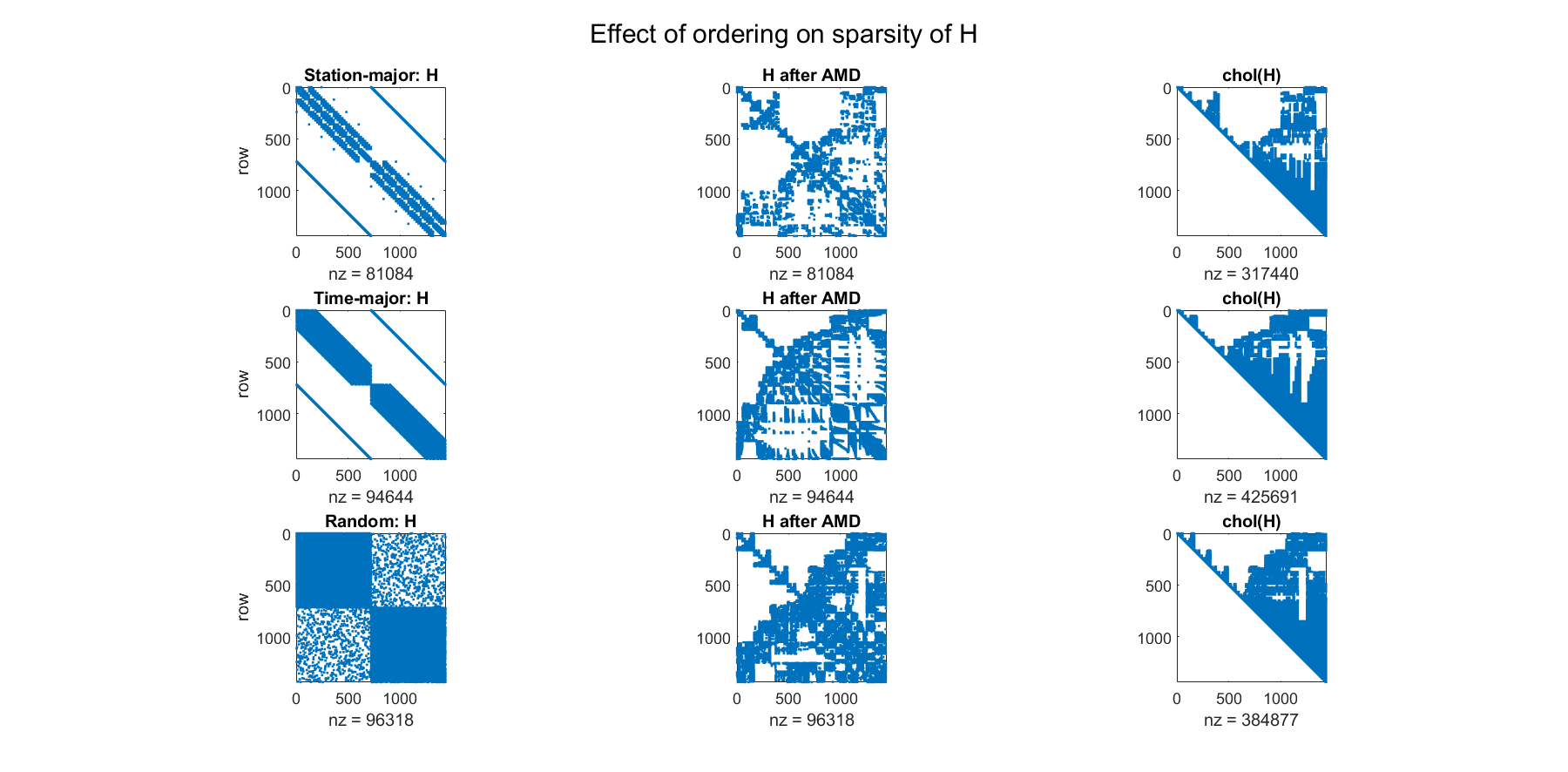}
\caption{%
Sparsity patterns of the precision matrix $\bm{H}$ under different
orderings.
Rows correspond to space-major ordering (used in the paper), time-major
ordering, and random ordering.
Columns show (left) $\bm{H}$ in its Nearest-Neighbor, (center)
$\bm{H}$ after a fill-reducing approximate minimum degree (AMD)
permutation, and (right) the sparse Cholesky factor.
While the apparent bandedness of $\bm{H}$ depends strongly on the
ordering, $\bm{H}$ remains sparse and admits efficient sparse Cholesky
factorization under all orderings considered.}
\label{fig:H_ordering}
\end{figure}

Figure~\ref{fig:H_ordering} visualizes the sparsity patterns corresponding to
three representative orderings.
The left column shows $\bm{H}$ in its Nearest-Neighbor, illustrating that
the apparent bandedness depends strongly on how the latent vector is ordered.
The center and right columns show $\bm{H}$ after AMD permutation and its
sparse Cholesky factor, respectively.
Although visual bandedness is not invariant to ordering, the AMD permutation
substantially stabilizes the sparsity structure relevant for computation.
Accordingly, likelihood evaluation and inference in all experiments rely on
sparse Cholesky factorization with a fill-reducing permutation, rather than
on any assumed intrinsic banded structure of $\bm{H}$.

\begin{table}[t]
\centering
\caption{Effect of ordering on sparsity of the precision matrix $\bm{H}$
for a simulated space--time data set with $n_L=n_H=720$.
Reported are the number of nonzeros in $\bm{H}$ and in the sparse
Cholesky factor after an approximate minimum degree (AMD) permutation.}
\label{tab:H_ordering}
\begin{tabular}{lccc}
\toprule
Ordering & nnz($\bm{H}$) & Density & nnz(chol($\bm{H}$)) \\
\midrule
Time-major           & 94{,}644 & 0.0456 & 425{,}691 \\
Space-major          & 81{,}084 & 0.0391 & 317{,}440 \\
Random               & 96{,}318 & 0.0464 & 384{,}877 \\
\bottomrule
\end{tabular}
\end{table}

\subsection{Effect of ordering strategy on likelihood accuracy}
\label{sec:supp_ordering_accuracy}

Table \ref{OrderingStrategies} depicts the averages of small montecarlo experiment of 20 run that test different ordering strategies. We iused the following metrics:
\begin{align}
\text{diffAbs} &= \lvert \text{NLML}_V - \text{NLML}_E \rvert, \\
\text{diffRel} &= \frac{\text{diffAbs}}{\max(\lvert \text{NLML}_E \rvert, 10^{-12})},
\end{align}
where the $\text{NLML}_V$ is the likelihood computed with the Vecchia approach while $\text{NLML}_E$ is the exact likelihood.
The results show a clear and consistent trend across all ordering strategies: increasing the number of neighbours' elements ($nn$) systematically improves the quality of the Vecchia approximation, as reflected by the monotonic decrease in both absolute and relative error metrics. This improvement comes at the expected cost of increased fill-in, as measured by $\mathrm{nnz}(R)$, confirming the standard accuracy--sparsity trade-off inherent to Vecchia-type approximations.

For small conditioning sets ($nn <15$), ordering plays a substantial role. The space-major strategy yields the lowest approximation error, closely followed by Random ordering, indicating that spatial locality dominates the dependence structure when only a limited number of neighbors is used. In contrast, time-major and time-major + RandSpace perform worse in this regime, suggesting that purely temporal prioritization does not capture the strongest short-range dependencies when the conditioning sets are small.

As $nn$ increases, however, Random ordering achieving the smallest relative error at $nn = 40$. Notably, the space-major strategy exhibits an early saturation effect, with limited improvement beyond $nn = 30$, whereas the other orderings continue to benefit from larger conditioning sets.

From a computational perspective, time-major ordering produces substantially sparser factors at small $nn$, yielding much smaller $\mathrm{nnz}(R)$ compared to the other strategies. This indicates a favorable sparsity--accuracy trade-off when computational efficiency is a priority. Conversely, strategies that incorporate stronger spatial mixing achieve slightly better accuracy at the expense of increased fill-in.

Overall, the results suggest that spatial structure is the dominant driver of local dependence in this setting, while temporal ordering primarily affects sparsity patterns. With sufficiently large conditioning sets, the choice of ordering becomes less critical, as the approximation error converges across strategies.

\begin{table}[ht]
\centering
\caption{Vecchia approximation results grouped by ordering strategy and increasing number of neighbors ($nn$). Values reported as mean $\pm$ standard deviation. Results computed by averaging 20 independent run.}
\small
\begin{tabular}{llccc}
\toprule
Ordering & $nn$ & DiffAbs & DiffRel & nnz($R$) \\
\midrule

\multirow{5}{*}{Random / Random}
& 10 & 649.3 $\pm$ 238.77 & 0.05257 $\pm$ 0.01695 & $6.78{\times}10^5 \pm 14675$ \\
& 15 & 372.21 $\pm$ 151.86 & 0.02993 $\pm$ 0.01062 & $7.98{\times}10^5 \pm 13781$ \\
& 20 & 209.59 $\pm$ 86.64 & 0.01679 $\pm$ 0.00577 & $8.97{\times}10^5 \pm 14032$ \\
& 30 & 78.38 $\pm$ 44.73 & 0.00627 $\pm$ 0.00341 & $9.55{\times}10^5 \pm 13320$ \\
& 40 & 49.80 $\pm$ 31.06 & 0.00395 $\pm$ 0.00209 & $9.71{\times}10^5 \pm 16826$ \\
\midrule

\multirow{5}{*}{space-major}
& 10 & 633.47 $\pm$ 213.66 & 0.05154 $\pm$ 0.01699 & $6.61{\times}10^5$ \\
& 15 & 397.40 $\pm$ 162.91 & 0.03274 $\pm$ 0.01394 & $7.90{\times}10^5$ \\
& 20 & 275.90 $\pm$ 151.80 & 0.02284 $\pm$ 0.01247 & $8.66{\times}10^5$ \\
& 30 & 211.86 $\pm$ 113.41 & 0.01676 $\pm$ 0.00723 & $9.34{\times}10^5$ \\
& 40 & 208.23 $\pm$ 90.59 & 0.01672 $\pm$ 0.00674 & $9.56{\times}10^5$ \\
\midrule

\multirow{5}{*}{time-major + RandSpace}
& 10 & 847.84 $\pm$ 319.27 & 0.06879 $\pm$ 0.02362 & $7.12{\times}10^5 \pm 15392$ \\
& 15 & 662.50 $\pm$ 274.92 & 0.05358 $\pm$ 0.02022 & $8.81{\times}10^5 \pm 9542$ \\
& 20 & 465.11 $\pm$ 162.65 & 0.03771 $\pm$ 0.01238 & $9.65{\times}10^5 \pm 9809$ \\
& 30 & 225.33 $\pm$ 99.84 & 0.01811 $\pm$ 0.00720 & $9.88{\times}10^5 \pm 9917$ \\
& 40 & 63.41 $\pm$ 38.62 & 0.00513 $\pm$ 0.00298 & $9.96{\times}10^5 \pm 11056$ \\
\midrule

\multirow{5}{*}{time-major}
& 10 & 723.65 $\pm$ 312.90 & 0.05850 $\pm$ 0.02351 & $1.12{\times}10^5$ \\
& 15 & 538.02 $\pm$ 152.57 & 0.04390 $\pm$ 0.01246 & $4.68{\times}10^5$ \\
& 20 & 412.86 $\pm$ 144.59 & 0.03352 $\pm$ 0.01162 & $6.89{\times}10^5$ \\
& 30 & 212.16 $\pm$ 97.58 & 0.01720 $\pm$ 0.00741 & $7.78{\times}10^5$ \\
& 40 & 50.20 $\pm$ 43.83 & 0.00411 $\pm$ 0.00354 & $8.06{\times}10^5$ \\
\bottomrule
\end{tabular}
\end{table}\label{OrderingStrategies}

\section{Generalised least squares mean removal}\label{app:gls_mean}

In multi-fidelity spatio-temporal Gaussian process  models, low-fidelity and high-fidelity  observations often share a common latent structure but differ by systematic baseline shifts. These shifts arise from sensor calibration differences, aggregation effects, or unresolved physics, and manifest as different intercept (offset) terms across fidelities.

To account for this, we incorporate a Generalised Least Squares mean model with fidelity-specific offsets prior to covariance-based inference. In other words, a GLS procedure with intercepts for each fidelity level.

Let
\[
\bm{y} =
\begin{bmatrix}
\bm{y}_L \\
\bm{y}_H
\end{bmatrix}
\in \mathbb{R}^{n_L + n_H},
\]
denote the stacked LF and HF observations, and let $\bm{K}$ denote the joint covariance implied by the multi-fidelity GP model. We define a mean function of the form
\[
\mathbb{E}[\bm{y}] = \bm{G}\boldsymbol{\beta},
\]
where $\boldsymbol{\beta} = (\beta_L, \beta_H)^\top$ contains separate intercepts for LF and HF, and the design matrix $\bm{G} \in \mathbb{R}^{(n_L+n_H)\times 2}$ is given by
\[
\bm{G} =
\begin{bmatrix}
\bm{1}_{N_L} & \bm{0}_{N_L} \\
\bm{0}_{N_H} & \bm{1}_{N_H}
\end{bmatrix}.
\]

This structure allows each fidelity to have its own baseline level while sharing a common covariance structure for residual variation.

\subsection*{GLS estimation of offsets}

Given the covariance $\bm{K}$, the GLS estimator of the intercept coefficients is
\[
\hat{\boldsymbol{\beta}}
=
(\bm{G}^\top \bm{K}^{-1} \bm{G})^{-1}
\bm{G}^\top \bm{K}^{-1} \bm{y}.
\]

The corresponding residual vector is
\[
\tilde{\bm{y}}
=
\bm{y} - \bm{G}\hat{\boldsymbol{\beta}},
\]
which is used in the marginal likelihood and all subsequent covariance-based computations.

\subsection*{Adaptive GLS with Spatial Trends}

To account for further systematic discrepancies between fidelities that may vary across the study area, we implement an \textit{adaptive} GLS procedure. Unlike the standard model which assumes a single global offset for each fidelity, the adaptive version incorporates a spatial linear trend into the design matrix $\bm{G}_{gls}$.

In this configuration, the mean for each fidelity $f \in \{L, H\}$ is modeled as a function of the spatial coordinates $\bm{s} = (s_{lat}, s_{lon})^\top$:
\[
\mathbb{E}[y_f(\bm{s})] = \beta_{f,0} + \beta_{f,1} s_{lat} + \beta_{f,2} s_{lon}.
\]

The resulting design matrix $\bm{G}_{gls} \in \mathbb{R}^{(n_L+n_H) \times 6}$ is constructed as a block-diagonal matrix:
\[
\bm{G}_{gls} = \begin{bmatrix} \bm{G}_L & \bm{0} \\ \bm{0} & \bm{G}_H \end{bmatrix}, \quad \text{where} \quad \bm{G}_f = [\bm{1}, \bm{s}_{f,lat}, \bm{s}_{f,lon}].
\]

The coefficients $\boldsymbol{\beta} = (\beta_{L,0}, \beta_{L,1}, \beta_{L,2}, \beta_{H,0}, \beta_{H,1}, \beta_{H,2})^\top$ are estimated using the joint covariance $\bm{K}$ via the Generalized Least Squares estimator:
\[
\hat{\boldsymbol{\beta}} = (\bm{G}_{gls}^\top \bm{K}^{-1} \bm{G}_{gls})^{-1} \bm{G}_{gls}^\top \bm{K}^{-1} \bm{y}.
\]

This approach allows the model to "adapt" to regional shifts—for instance, if the high-fidelity sensor exhibits a different bias in the north of the domain compared to the south. By capturing these linear spatial trends in the mean function, the Gaussian process residuals $\tilde{\bm{y}} = \bm{y} - \bm{G}_{gls}\hat{\boldsymbol{\beta}}$ are better centered, improving the stability of the Vecchia approximation and the accuracy of the cross-fidelity scaling $\rho(\bm{s})$.

\subsection*{Role in likelihood and prediction}

Removing the GLS mean before evaluating the likelihood ensures that the covariance model is fit to centred residuals rather than absorbing systematic level differences between fidelities. This separation of mean and covariance has several important consequences:
 It prevents the cross-fidelity scaling function (e.g., $\rho(\cdot)$) from compensating for simple baseline shifts. It improves numerical stability of the precision matrix factorisation, particularly in large Vecchia-approximated systems. It yields calibrated predictive variances, since uncertainty is attributed to stochastic structure rather than deterministic offsets. For prediction at new HF locations $\bm{x}_\ast$, the latent GP prediction is computed using $\tilde{\bm{y}}$, and the HF intercept $\hat{\beta}_H$ is added back to obtain the final predictive mean:
\[
\mu_H(\bm{x}_\ast) = \hat{\beta}_H + \mu_0(\bm{x}_\ast),
\]
where $\mu_0(\bm{x}_\ast)$ denotes the zero-mean GP predictor.

\begin{figure}
    \centering
    \includegraphics[width=1\linewidth]{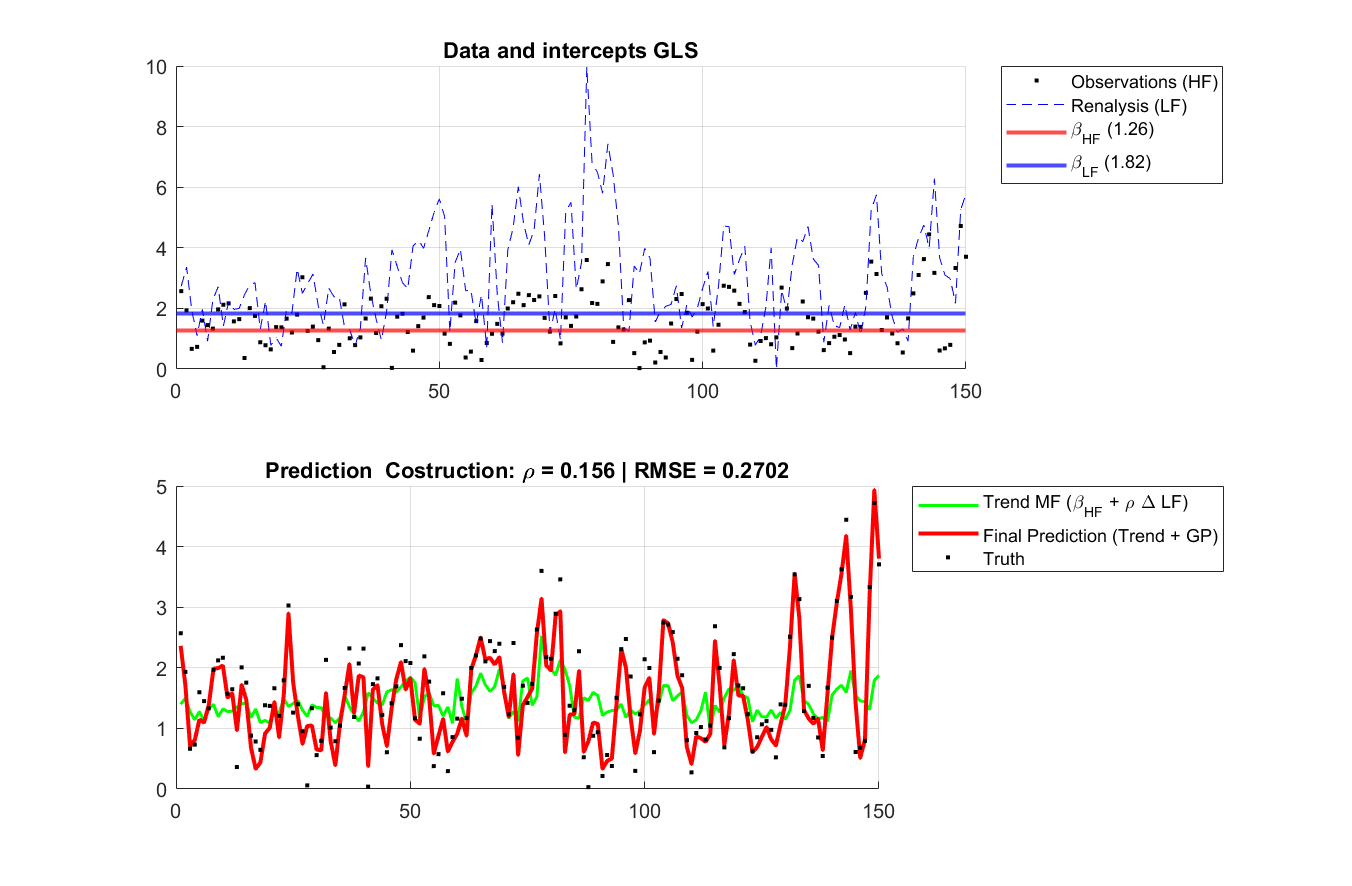}
    \caption{The first panel illustrates the estimated GLS intercepts ($\hat{\beta}_L, \hat{\beta}_H$), representing the baseline levels for each fidelity. The second panel demonstrates the decomposition of the multi-fidelity signal: the green line tracks the systematic offset (the mean model), while the gap between the green and red lines represents the residual GP process accounting for shared spatio-temporal variability.}
    \label{fig:GLS}
\end{figure}

The GLS mean adjustment can be interpreted as explicitly modelling
fidelity-specific systematic offsets, while allowing the Gaussian
process component to focus exclusively on learning the shared
spatio-temporal structure and the residual variability.
In this way, large-scale mean discrepancies between low- and
high-fidelity sources are separated from the correlation structure,
leading to a cleaner decomposition of variability.
Empirically, this separation substantially improves both predictive
accuracy and uncertainty quantification in multi-fidelity settings,
as the covariance hyperparameters are no longer forced to compensate
for mean mis-specification.
Figure~\ref{fig:GLS} illustrates how the GLS component contributes
to the construction of the final prediction.
The first panel displays the two data sources—high fidelity (dots)
and low fidelity (blue dotted line)—together with their respective
mean offset terms.
The second panel shows the formation of the trend component
$\beta_{\mathrm{HF}} + \rho \,\Delta_{\mathrm{LF}}$ (green curve),
and the final prediction obtained after adding the discrepancy
process (red curve).

\begin{figure}
    \centering
    \includegraphics[width=1\linewidth]{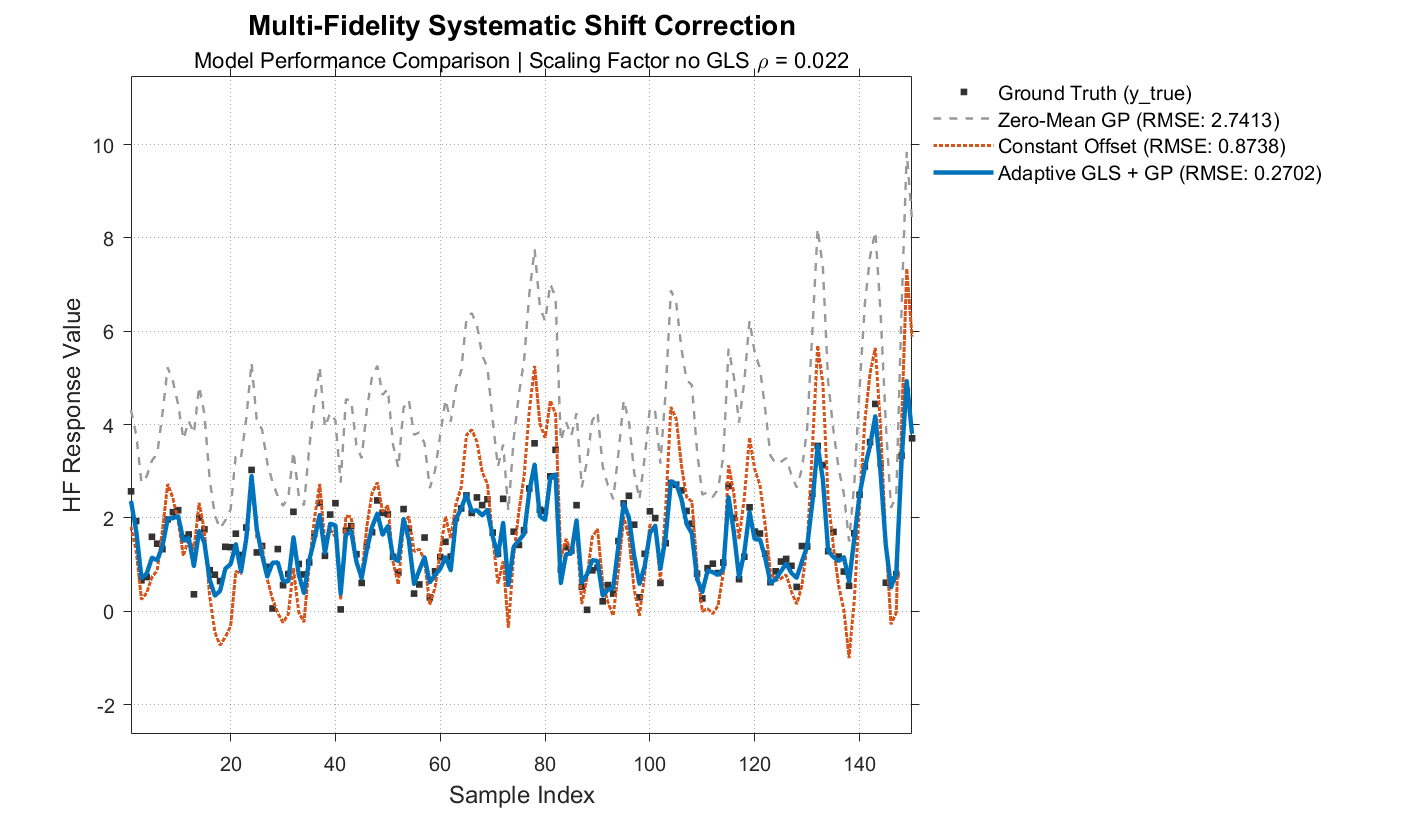}
    \caption{Comparison of multi-fidelity GP prediction schemes: The proposed Adaptive GLS model (blue) effectively captures the systematic shift between fidelities, outperforming the zero-mean baseline (grey) and the fixed constant-mean model (red). While the baseline models exhibit significant bias due to uncorrected offsets, the GLS approach dynamically aligns the mean structure, resulting in superior predictive accuracy as indicated by the comparative RMSE values.}
    \label{fig:NaturalOffset}
\end{figure}

Figure~\ref{fig:NaturalOffset} compares three modelling strategies:
(i) a multi-fidelity Gaussian process without mean adjustment,
(ii) a generalized least squares model with a constant offset,
and (iii) a generalized least squares model with adaptive, location-specific offsets.

The red curve corresponds to the model with a constant offset,
where one offset parameter is estimated for the high-fidelity data
and one for the low-fidelity data, shared across all spatial locations.
The dotted grey curve represents the zero-mean multi-fidelity Gaussian process,
while the blue curve shows the model with adaptive offsets,
where each spatial location is allowed to have its own offset term.

In this particular example, the zero-mean model drives the
correlation parameter $\rho$ close to zero.
When $\rho = 0$, the coupling between the two fidelities disappears,
and the model effectively reduces to a single Gaussian process,
with all variation absorbed into the discrepancy component.

Introducing a constant offset improves the behaviour of the model,
but in this specific case it produces unrealistic (negative)
predictions.
Allowing for adaptive offsets provides a more flexible and stable
decomposition, preserving the multifidelity structure while preventing
the mean mis-specification from distorting the correlation parameter.

\paragraph{Simulation supplements}

\section{Simulation of spatio-temporal multi-fidelity data}\label{sec:sim_data}
We generate synthetic spatio-temporal multi-fidelity observations following the autoregressive
multi-fidelity relationship introduced in Section~\ref{MFmodel},
\begin{equation}\label{eq:sim_mf_relation}
    y_L(x) = f_L(x) + \varepsilon_L(x), 
    \qquad
    y_H(x) = f_H(x) = \rho\, y_L(x) + \delta(x),
\end{equation}
where $x=(s,t)$ denotes a spatio-temporal coordinate with $s=(s_1,s_2)$ and $t\in[0,1]$.
In the implementation, the LF field is generated as a latent Gaussian process plus an independent
measurement error, while the discrepancy is generated as an independent Gaussian process plus an
additional independent noise term. This construction yields an HF field that is correlated with the LF
field via $\rho$ and departs from it through $\delta$.

\paragraph{Spatio-temporal grid and indexing.}
We consider a regular spatial grid with $n_{\text{space}}\times n_{\text{space}}$ stations, hence
$N_s=n_{\text{space}}^2$ spatial locations $s_j=(s_{1j},s_{2j})$, $j=1,\dots,N_s$.
We also consider $N_t=n_{\text{time}}$ time points $t_k\in[0,1]$, $k=1,\dots,N_t$.
The full set of spatio-temporal inputs is
\[
x_{jk}=(s_j,t_k), \qquad j=1,\dots,N_s,\ \ k=1,\dots,N_t,
\]
and we stack the process values into vectors of length $N=N_sN_t$ using a consistent ordering
(e.g., time-fastest within station), matching the construction used later for the covariance matrices.

\paragraph{RBF kernels and target correlations.}
Both LF and discrepancy components use separable squared-exponential (RBF) covariance models.
For a generic component ``$\star\in\{L,\delta\}$'', we define spatial and temporal kernels
\begin{align}
k^{\star}_s(s,s') &= \sigma^2_{\star}\,
\exp\!\left(-\frac{1}{2}\sum_{d=1}^{2}\frac{(s_d-s'_d)^2}{\ell^{\star}_{s,d}{}^{2}}\right), \\
k^{\star}_t(t,t') &= \sigma^2_{\star}\,
\exp\!\left(-\frac{(t-t')^2}{2\,\ell^{\star}_t{}^{2}}\right),
\end{align}
with signal variances $\sigma^2_L$ and $\sigma^2_\delta$ and length-scales
$\ell^{\star}_{s,1},\ell^{\star}_{s,2},\ell^{\star}_t$.
Rather than specifying length-scales directly, we parameterize them via \emph{target}
nearest-neighbour correlations. Let $\Delta_t=1/(N_t-1)$ be the time spacing and let $d_s=1$ be the
unit spatial neighbour distance on the grid. For a desired correlation level $c\in(0,1)$ at distance $d$,
the RBF correlation satisfies $c=\exp\!\big(-\tfrac{1}{2}(d/\ell)^2\big)$, hence
\begin{equation}\label{eq:ell_from_corr}
\ell(c,d)= \frac{d}{\sqrt{-2\log(c)}}.
\end{equation}
Using \eqref{eq:ell_from_corr}, we set
\[
\ell^{L}_{s,1}=\ell^{L}_{s,2}=\ell(\texttt{target\_corr\_spaceL},d_s),\quad
\ell^{\delta}_{s,1}=\ell^{\delta}_{s,2}=\ell(\texttt{target\_corr\_spaceD},d_s),
\]
and similarly for time,
\[
\ell^{L}_{t}=\ell(\texttt{target\_corr\_time},\Delta_t),\quad
\ell^{\delta}_{t}=\ell(\texttt{target\_corr\_time},\Delta_t).
\]

\paragraph{Separable spatio-temporal covariance construction.}
Let $\bm{K}^{\star}_s\in\mathbb{R}^{N_s\times N_s}$ and
$\bm{K}^{\star}_t\in\mathbb{R}^{N_t\times N_t}$ be the spatial and temporal covariance matrices obtained
by evaluating $k_s^\star$ and $k_t^\star$ on $\{s_j\}$ and $\{t_k\}$, respectively.
We form a separable spatio-temporal covariance on the stacked vector using a Hadamard--Kronecker form,
equivalent to the standard separable model $\bm{K}^\star=\bm{K}^\star_s\otimes \bm{K}^\star_t$ up to
the chosen stacking convention:
\begin{equation}
\bm{K}^\star_{\text{full}}
=
\big(\bm{K}^\star_s \otimes \bm{1}_{N_t}\bm{1}_{N_t}^\top\big)
\odot
\big(\bm{1}_{N_s}\bm{1}_{N_s}^\top \otimes \bm{K}^\star_t\big)
\;\in\;\mathbb{R}^{N\times N},
\end{equation}
where $\otimes$ denotes the Kronecker product and $\odot$ the elementwise (Hadamard) product.

\paragraph{Latent LF field and LF observations.}
We draw the LF latent process on the full grid as
\begin{equation}
\bm{d}_L \sim \mathcal{N}\!\left(\bm{0},\ \bm{K}^{L}_{\text{full}} + \eta\bm{I}\right),
\end{equation}
where $\eta$ is a small jitter term added for numerical stability.
We then generate LF observations by adding i.i.d.\ measurement noise,
\begin{equation}
\bm{e}_L \sim \mathcal{N}\!\left(\bm{0},\ \sigma^2_{\text{noise},L}\bm{I}\right),
\qquad
\bm{f}_L = \bm{d}_L + \bm{e}_L,
\end{equation}
so that the LF output stored in the tables corresponds to $\bm{f}_L$.

\paragraph{Discrepancy field and HF observations.}
Independently, we draw a discrepancy component on the same spatio-temporal grid:
\begin{equation}
\bm{d}_\delta \sim \mathcal{N}\!\left(\bm{0},\ \bm{K}^{\delta}_{\text{full}} + \eta\bm{I}\right),
\qquad
\bm{e}_\delta \sim \mathcal{N}\!\left(\bm{0},\ \sigma^2_{\text{noise},\delta}\bm{I}\right),
\qquad
\bm{\delta}=\bm{d}_\delta+\bm{e}_\delta.
\end{equation}
Finally, the HF field is constructed through the autoregressive multi-fidelity relationship
\begin{equation}\label{eq:sim_hf}
\bm{f}_H = \rho\,\bm{f}_L + \bm{\delta}.
\end{equation}
This matches the modelling assumption in Section~\ref{MFmodel}: $\rho$ controls the linear
dependence between fidelities, while $\delta$ introduces structured spatio-temporal departures.

\paragraph{Train/test split by station.}
To mimic a spatially sparse HF deployment, we split the HF observations by station (not by time):
a fraction \texttt{train\_fraction} of stations is sampled uniformly at random as HF training stations,
and the remaining stations define the HF test set. All time points at a training station are included
in the HF training set, and all time points at test stations are held out. The LF table is retained over
all spatio-temporal points.
The random seed ensures reproducibility of both the field generation and the station split.

\section{Positive semidefiniteness of the spatially varying rescaling model}
\label{app:PSD_rho}

In this section we formally establish that the proposed multi-fidelity
Gaussian process model with a spatially varying rescaling function
$\rho(\cdot)$ induces a valid covariance function, i.e., a symmetric
positive semidefinite (PSD) covariance matrix.

\subsection{Model specification and hierarchical representation} 
Let $w_L \sim GP(0,k_L)$ and $w_\delta \sim GP(0,k_\delta)$ be independent
Gaussian processes representing the low-fidelity (LF) latent process and
the discrepancy process, respectively. Observations are generated according
to
\begin{align}
\bm{y}_L &= \bm{Z}_1 \bm{w}_L + \bm{\varepsilon}_L, \\
\bm{y}_H &= \bm{R}\bm{Z}_{21}\bm{w}_L + \bm{w}_\delta + \bm{\varepsilon}_\delta,
\end{align}
where $\bm{Z}_1$ and $\bm{Z}_{21}$ are fixed incidence matrices,
$\bm{R}=\mathrm{diag}(\rho(s_{H,1}),\ldots,\rho(s_{H,n_H}))$ is a diagonal
matrix containing evaluations of the spatially varying rescaling function at
the HF locations, and $\bm{\varepsilon}_L$, $\bm{\varepsilon}_\delta$ are
independent Gaussian noise terms with covariance $\bm{D}_\epsilon$.
All components are assumed mutually independent.

Defining the stacked latent vector
\[
\bm{w} =
\begin{pmatrix}
\bm{w}_L \\
\bm{w}_\delta
\end{pmatrix},
\qquad
\boldsymbol{\Sigma}_w =
\mathrm{blkdiag}(\boldsymbol{\Sigma}_L,\boldsymbol{\Sigma}_\delta),
\]
and the linear mapping
\[
\bm{A} =
\begin{pmatrix}
\bm{Z}_1 & \bm{0} \\
\bm{R}\bm{Z}_{21} & \bm{I}
\end{pmatrix},
\]
the joint observation vector $\bm{y}=(\bm{y}_L^\top,\bm{y}_H^\top)^\top$
can be written as
\[
\bm{y} = \bm{A}\bm{w} + \bm{\varepsilon}.
\]

---

\subsection{Implied covariance structure}
By standard properties of linear transformations of random vectors, the
covariance matrix of $\bm{y}$ is given by
\begin{equation}
\label{eq:K_factorization}
\bm{K}
=
\mathrm{Cov}(\bm{y})
=
\bm{A}\boldsymbol{\Sigma}_w\bm{A}^\top
+
\bm{D}_\epsilon.
\end{equation}

At the kernel level, this corresponds to the block structure
\begin{align}
k_{LL}(x,x') &= k_L(x,x'), \\
k_{HL}(x,x') &= \rho(s)\,k_L(x,x'), \\
k_{HH}(x,x') &= \rho(s)\rho(s')\,k_L(x,x') + k_\delta(x,x').
\end{align}
The HF--HF block therefore takes the PSD-preserving form
$\rho(s)\rho(s')k_L(x,x')$ plus an independent discrepancy term.

Although $\rho(\cdot)$ may vary spatially, it enters the model only through
a deterministic, location-dependent linear operator acting on the latent
process. In particular, letting $\bm{R}=\mathrm{diag}(\rho(s_{H,i}))$,
the mapping $\bm{w}_L \mapsto \bm{R}\bm{Z}_{21}\bm{w}_L$ is linear,
since for any $\bm{w}_L^{(1)},\bm{w}_L^{(2)}$ and scalars $a,b\in\mathbb{R}$,
\[
\bm{R}\bm{Z}_{21}\bigl(a\,\bm{w}_L^{(1)} + b\,\bm{w}_L^{(2)}\bigr)
=
a\,\bm{R}\bm{Z}_{21}\bm{w}_L^{(1)}
+
b\,\bm{R}\bm{Z}_{21}\bm{w}_L^{(2)}.
\]
Consequently, the covariance contribution
$\bm{R}\bm{Z}_{21}\boldsymbol{\Sigma}_L\bm{Z}_{21}^\top\bm{R}$
is a congruence transformation of a positive semidefinite matrix and is
therefore itself positive semidefinite.


\section{Additional results}

This section provides supplementary material referenced in the main text. Figure F.1 illustrates the computational efficiency of the proposed method, while Table B.2 presents a comparative analysis of the accuracy achieved across different ordering strategies.

\clearpage   

\clearpage   

\begin{figure}
    \centering
    \includegraphics[width=1\linewidth]{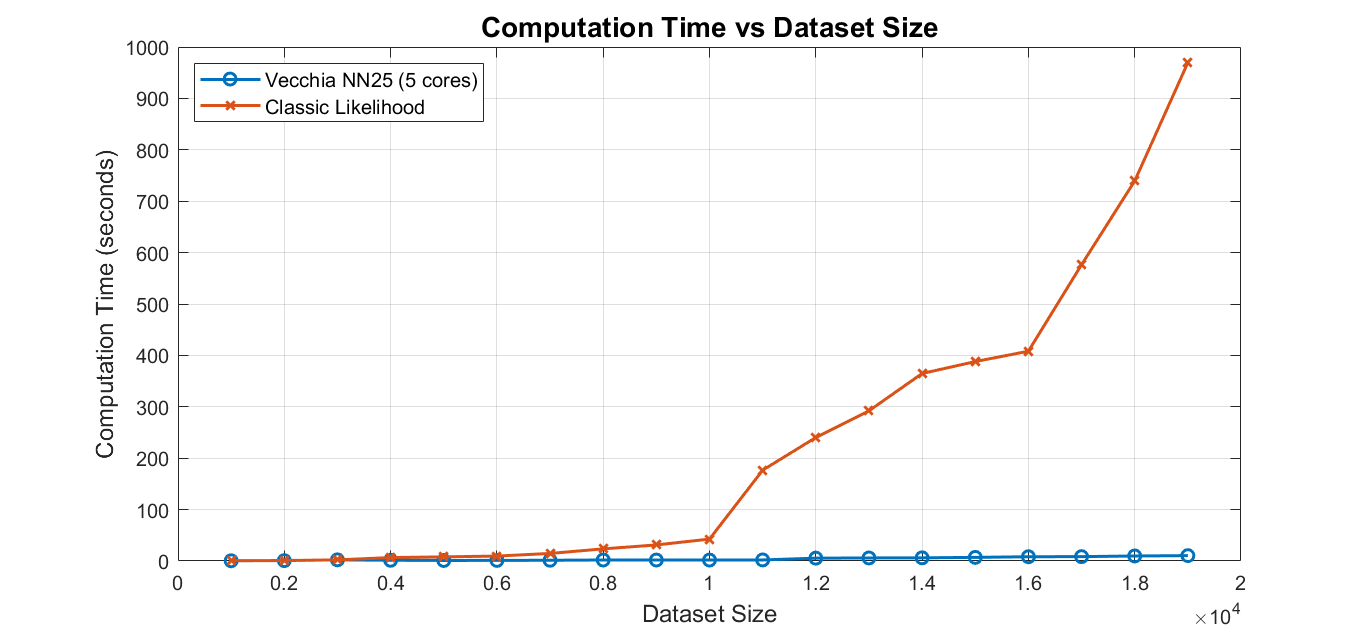}
    \caption{Example of the likelihood computation time by increasing dataset size.}
    \label{fig:computation_time}
\end{figure}

\end{document}